\documentclass[onecolumn,draftclsnofoot,journal]{IEEEtran} %
\usepackage{amsmath,amssymb,amsfonts}
\usepackage{algorithmic}
\usepackage{mathtools}
\usepackage{algorithm}
\usepackage{array}
\usepackage[font=normalsize]{subfig}
\usepackage{textcomp}
\usepackage{stfloats}
\usepackage{url}
\usepackage{verbatim}
\usepackage{siunitx}
\usepackage{graphicx}
\usepackage{cite}
\usepackage{makecell}
\usepackage[switch]{lineno} %
\usepackage{booktabs}

\usepackage[dvipsnames]{xcolor}
\usepackage{subfig}
\usepackage{placeins}
\usepackage{xifthen}
\usepackage{colortbl}
\usepackage{array}
\newcolumntype{H}{>{\setbox0=\hbox\bgroup}c<{\egroup}@{}}
\newcommand{\prob}[1][]{%
\ifthenelse{\isempty{#1}}%
      {\ensuremath{P}}%
    {\ensuremath{P\left\(#1\right\)}}%
}
\newcommand{\vect}[1]{\boldsymbol{\mathrm{#1}}}
\newcommand{\mat}[1]{\boldsymbol{\mathrm{#1}}}

\newcommand{\diag}[1]{\mathrm{diag}\left(#1\right)}

\newcommand{\expt}[1]{\mathbb{E}\left(#1\right)}

\usepackage{pgfplots}
\usepackage{nicefrac}
\usepgfplotslibrary{groupplots,dateplot}

\pgfplotsset{compat=1.18}
\usepgfplotslibrary{fillbetween}
\usepackage{tikz}
\usetikzlibrary{positioning}
\usepackage{hyperref}
\usepackage{acronym}
\usepackage[capitalize]{cleveref}
\usepackage{subfig}
\usetikzlibrary{patterns}
\usepgfplotslibrary{polar}
\DeclareMathOperator*{\argmin}{arg\,min}

\usetikzlibrary{shapes}

\usepackage[acronym]{glossaries}
\newacronym[plural=BSs,firstplural=base stations (BSs)]{bs}{BS}{base station}
\newacronym[plural=PAs,firstplural=power amplifiers (PAs)]{pa}{PA}{power amplifier}
\newacronym{sdr}{SDR}{signal-to-distortion ratio}
\newacronym{snr}{SNR}{signal-to-noise ratio}
\newacronym{sndr}{SNDR}{signal-to-noise-and-distortion ratio}
\newacronym{snidr}{SNIDR}{signal-to-noise-and-interference-and-distortion ratio}
\newacronym{los}{LOS}{line-of-sight}
\newacronym{z3ro}{Z3RO}{zero third-order distortion}
\newacronym{mimo}{MIMO}{multiple input multiple output}
\newacronym{mrt}{MRT}{maximum ratio transmission}
\newacronym{dpd}{DPD}{digital pre-distortion}

\newacronym{ula}{ULA}{uniform linear array}

\newacronym{psd}{PSD}{power spectral density}
\newacronym{aclr}{ACLR}{adjacent channel leakage ratio}
\newacronym{cpu}{CPU}{central processing unit}
\newacronym{iid}{i.i.d.}{independently and identically distributed}
\newacronym{ue}{UE}{user equipment}
\newacronym{nlos}{NLoS}{non-line-of-sight}
\newacronym{hpbm}{HPBM}{half power beam width}
\newacronym{rts}{RTS}{ray tracing simulator}
\newacronym{ag}{AG}{array gain}
\newacronym{arp}{ARP}{Antenna Reference Point}
\newacronym{ecdf}{eCDF}{empirical cumulative distribution function}
\newacronym{evm}{EVM}{Error Vector Magnitude}
\newacronym{fft}{FFT}{fast Fourier transform}
\newacronym{ib}{IB}{in-band}
\newacronym{im}{IM}{intermodulation}
\newacronym{mf}{MF}{matched filtering}
\newacronym{mr}{MR}{maximum ratio}
\newacronym{ofdm}{OFDM}{orthogonal frequency division multiplexing}
\newacronym{oob}{OOB}{out-of-band}
\newacronym{papr}{PAPR}{peak-to-average power ratio}
\newacronym{rx}{RX}{Receiver}
\newacronym{trp}{TRP}{Transmission Reception Point}
\newacronym{tx}{TX}{transmitter}
\newacronym{zf}{ZF}{zero forcing}
\newacronym{dl}{DL}{downlink}
\newacronym{rzf}{RZF}{regularized zero forcing}
\newacronym{slc}{SLC}{spatial leakage suppression}
\newacronym{kpi}{KPI}{key performance indicator}
\newacronym{awgn}{AWGN}{additive white Gaussian noise}
\newacronym{qam}{QAM}{quadrature amplitude modulation}
\newacronym{amam}{AM/AM}{amplitude modulation to amplitude modulation}
\newacronym{ampm}{AM/PM}{amplitude modulation to phase modulation}
\newacronym{zmcscg}{ZMCSCG}{zero mean circularly symmetric complex Gaussian}

\newacronym{gnn}{GNN}{graph neural network}
\newacronym{ccnn}{CCNN}{circular convolutional neural network}
\newacronym{lrelu}{LReLU}{leaky rectified linear unit }
\newacronym{sdg}{SDG}{Sustainable Development Goal}
\newacronym{ibo}{IBO}{input back-off}
\newacronym{cdf}{CDF}{cumulative distribution function}
\newacronym{nn}{NN}{neural network}
\newacronym{mlp}{MLP}{multilayer perceptron}
\newacronym{mmimo}{mMIMO}{massive MIMO}
\newacronym{dab}{DAB}{distortion-aware beamforming}
\newacronym{flops}{FLOPs}{floating point operations}
\newacronym{mmwave}{mmWave}{millimeter-wave}
\newacronym{gops}{GFLOPs/s}{giga floating point operations per second}
\newacronym{sp}{SP}{single precision}
\newacronym{dsp}{DSP}{digital signal processing}
\glsdisablehyper

\usetikzlibrary{external}
\usepackage{amsthm}
\newtheoremstyle{mystyle}%
  {}%
  {}%
  {\itshape}%
  {}%
  {\bfseries}%
  {.}%
  { }%
  {}%
\theoremstyle{mystyle}

\begin{document}

\title{Toward Energy-Efficient Massive MIMO: Graph Neural Network Precoding for Mitigating Non-Linear PA Distortion
}

\author{Thomas Feys, Liesbet Van der Perre,
        François Rottenberg
\thanks{The authors are with ESAT-DRAMCO, Campus Ghent, KU Leuven, 9000 Ghent, Belgium (email: \href{mailto:thomas.feys@kuleuven.be}{thomas.feys@kuleuven.be}).}
\thanks{We would like to thank NVIDIA for providing the GPU that greatly accelerated our simulations.}}

\maketitle

\begin{abstract}
Massive MIMO systems are typically designed assuming linear \acrfullpl{pa}. However, PAs are most energy efficient close to saturation, where non-linear distortion arises. For conventional precoders, this distortion can coherently combine at user locations, limiting performance. We propose a \acrfull{gnn} to learn a mapping between channel and precoding matrices, which maximizes the sum rate affected by non-linear distortion, using a high-order polynomial PA model. In the distortion-limited regime, this \acrshort{gnn}-based precoder outperforms \acrfull{zf}, \acrshort{zf} plus \acrfull{dpd} and the \acrfull{dab} precoder from the state-of-the-art. At an input back-off of \SI{-3}{\decibel} the proposed precoder compared to \acrshort{zf} increases the sum rate by 8.60 and 8.84 bits/channel use for two and four users respectively. Radiation patterns show that these gains are achieved by transmitting the non-linear distortion in non-user directions. In the four user-case, for a fixed sum rate, the total consumed power (PA and processing) of the GNN-precoder is 3.24 and 1.44 times lower compared to \acrshort{zf} and \acrshort{zf} plus \acrshort{dpd} respectively. A complexity analysis shows six orders of magnitude reduction compared to \acrshort{dab} precoding. This opens perspectives to operate PAs closer to saturation, which drastically increases their energy efficiency.
\end{abstract}


\section{Introduction}
\label{sec:intro}
\subsection{Problem Statement}

\IEEEPARstart{T}{he} wireless communications sector has seen a continuous increase in its carbon footprint and electricity usage~\cite{trends2040, real_climate_impact_ict}, leading to concerns about meeting the targets set forth by Europe's Green Deal~\cite{greendeal}, the United Nations \gls{sdg}~\cite{sdgs} and the Paris Agreement~\cite{paris_agreement}. This growth in emissions is in stark contrast with today's climate ambitions, where emissions must shrink rather than grow. Next to this, energy consumption has become an important factor in the operating expenses of wireless networks, which is further expected to worsen. Both the exponential growth in data~\cite{ericsonmobility} and the increased price of energy, contribute to this. 

The \gls{bs} is responsible for the majority of the energy consumption of mobile networks~\cite{towards_green_5g}. Within the \gls{bs}, the \glspl{pa} are responsible for a major component of the energy consumption, which is typically $\sim50-80\%$~\cite{energy_performance_ericsson, how_much_energy, huawei_pa}. Operating the \glspl{pa} in an energy-efficient manner has been a significant challenge. \Glspl{pa} are most efficient close to their saturation point, where non-linear distortion arises~\cite{rf_imperfections_pas}. The non-linear distortion limits the signal quality, leading to lower \glspl{sndr}. This leads to a trade-off between energy efficiency and linearity. In the past, this trade-off between energy efficiency and linearity has been skewed towards linearity to maintain system capacity. This was done by operating \glspl{pa} at a considerable back-off power in order to stay in the linear regime, which is detrimental for the energy efficiency. As a result, the current technology's energy efficiency of \glspl{pa} is typically low, ranging from $5\%$ to $30\%$~\cite{how_much_energy}.

In this study, we present an approach to improve the energy efficiency of \glspl{pa} by operating them closer to saturation. To do so, we study how a neural network can learn a \gls{mmimo} precoder that maximizes the sum rate, in the presence of non-linear distortion. This approach achieves a high capacity while using less back-off, thereby improving the energy efficiency. This in turn can reduce the operating expenses of wireless networks. Moreover, the cost of the \glspl{pa} is an important component in the Bill-of-Materials (BOM) of wireless access equipment~\cite{energy_performance_ericsson}. This cost is highly related to the maximum output power. Hence, by operating the \glspl{pa} closer to saturation a lower maximum output power can be used to achieve the same average output power. The proposed method can be of particular interest for millimeter-wave systems which are typically highly non-linear~\cite{mmwave_nonlinearities, Eriksson2019NonlinearEO}. %

\subsection{State-of-the-Art}

The energy efficiency of the \gls{pa} is typically limited by the need for a large back-off to remain within its linear regime. This is becoming less preferable given the growing emphasis on reducing energy consumption. Efforts to linearize the \gls{pa} such as \gls{dpd} are used in practical systems~\cite{powerconsumption_dpd}. However, in \gls{mmimo} systems, \gls{dpd} has to be deployed at each antenna. Recent studies show that when 5G systems would use fully digital beamforming, a large number of antennas and a large bandwidth, these \glspl{dpd} would introduce a power consumption that is even higher than the \gls{pa} consumption itself\cite{powerconsumption_dpd}. Moreover, the performance of \gls{dpd} is limited by clipping, i.e., the \gls{pa} can only be linearized up to the saturation point, so that a relatively large back-off is still required, even for a perfect \gls{dpd}. \Gls{papr} reduction techniques have been studied, which can reduce the amount of back-off needed~\cite{papr_reduction}. For instance, in~\cite{constant_envelope} a constant-envelop precoder has been proposed. However, practical implementations encountered challenges due to the substantial complexity of digital signal processing.

Various other approaches to the distortion problem have been proposed. For instance, in~\cite{virtualdpd} a neural network-based digital predistorter is proposed. This \gls{dpd} is applied before the precoding stage which leads to a complexity that scales with the number of users, rather than the number of antennas. Similarly, in~\cite{autoprecoder} an end-to-end system is proposed where neural-network based precoding and decoding are jointly learned. In this scheme, the transmit symbol is essentially pre-distorted using a neural network, after which a zero-forcing precoder is applied, finally a neural network-based based decoder is deployed at the receiver side. Alternatively, in~\cite{jointprecodingandpowerallocation} joint precoding and power allocation is proposed to solve the sum rate maximization problem in the presence of non-linear \glspl{pa}. The authors solve the optimization problem using an iterative approach that alternates between optimizing the precoding and power allocation matrices. In a similar vein, in~\cite{power_allocation_sofie} a multiplier punitive method based power allocation algorithm is proposed which maximizes the spectral efficiency with awareness of the non-linear amplifiers. Additionally, in~\cite{cooperativebeamforming} the problem of non-linear \glspl{pa} in distributed networks is considered. The authors propose a constrained unsupervised learning approach to solve the sum rate maximization problem in the presence of non-linear \glspl{pa} with a particular focus on reducing the communication overhead between the distributed access points.

Another compelling direction aims at incorporating knowledge of the distortion into the precoder design\cite{z3ro, zerofamily, z3ro_val, dab_mmwave}. This allows for the spatial suppression of the distortion in the user directions, producing considerable gains over classical precoders. Moreover, this offers an advantage over \gls{dpd} based solutions as the distortion created by saturation based clipping can be radiated in non-user directions, which improves the signal quality. Unfortunately, these solutions are still limited in their practical implementation, either because of their computational complexity or in terms of simplifying assumptions that need to be addressed. More specifically, in~\cite{dab_mmwave}, the solution to the precoding problem is obtained by solving a non-convex optimization problem with a projected gradient descent-based procedure. As the problem is non-convex, the procedure is executed multiple times to obtain a close-to-optimal solution, which requires a high computational complexity. As an alternative solution for the problem, the authors in~\cite{z3ro} derived a globally optimal closed-form solution for the simplified single-user case, a third order \gls{pa} model and a \gls{los} channel, which was later extended to a general channel in~\cite{zerofamily}. There is thus a need for a solution that has low complexity and can address the challenging case of spatially multiplexing many users. Additionally, in order to work close to saturation, this solution should be designed for polynomial \gls{pa} models that have a higher order than the third order model, as the third order model is only viable far away from saturation.

\subsection{Contributions}

In this work, we propose the use of a \gls{gnn} to find a mapping from the channel matrix to the precoding matrix. This mapping is learned under the presence of non-linear \glspl{pa} operating close to their saturation point, which increases their energy efficiency but as a consequence introduces non-linear distortion. The need for machine learning arises from the non-linear and non-convex nature of the optimization problem, which limits classical linear signal processing solutions. By learning the mapping from channel matrix to precoding matrix, a lot of the complexity is offloaded to the training step, which reduces the online computational complexity. This allows for a practical low-complexity solution even for the multi-user precoding problem under the presence of high-order non-linear \gls{pa} distortion. Previous works have not addressed this problem with such a practical low-complexity solution. As such, this opens perspectives to operate \glspl{pa} closer to their saturation point, which drastically increases their energy efficiency.

More specifically, the structure of the paper and our contributions are as follows: 
\begin{itemize}
    \item ~\Cref{sec:problem_forulation} describes the system model along with the non-linear \gls{pa} models which are used to train and evaluate the proposed \gls{gnn} precoder. Next to this, the optimization problem considering high-order \glspl{pa} is formulated. It is outlined how this formulation is used as a loss function to train a neural network precoder.
    
    \item We compare a number of neural network architectures for linear precoding in~\cref{sec:nn_comp}. This is done under the assumption of a simplified third-order \gls{pa} model. A \gls{gnn} is shown to outperform others such as the \gls{ccnn} presented in our previous work~\cite{icc_ccnn}, where a~\gls{ccnn} was studied in the context of a simplified third-order \gls{pa} model. Based on this comparison the \gls{gnn} is selected for further training and evaluation on a high-order \gls{pa} model.

    \item In~\cref{sec:gnn_basic}, we characterize the \gls{gnn} architecture and how it can be adopted for precoding. This architecture is further extended in~\cref{sec:snrgnn} to include the \gls{snr} as an additional input, which allows the precoder to perform well over the full \gls{snr} range, rather than having to be retrained at each \gls{snr} point to reach the desired performance. 

    \item The complexity of the proposed \gls{gnn} precoder is analysed in~\cref{sec:complexity}. Additionally, the complexity of the benchmark \gls{dab} precoder from~\cite{dab_mmwave} has been computed and used to compare to the \gls{gnn} precoder.

    \item \cref{sec:results} provides extensive simulation results to validate the performance of the \gls{gnn}-based precoder under the presence of high-order non-linear \glspl{pa}. This includes the computation of the power consumption reduction, where the total consumed power (PA and processing) of the \gls{gnn} precoder is quantified.

    \item \Cref{sec:conclusion} concludes the paper.

\end{itemize}

\subsection*{Notations}
 Vectors and matrices are denoted by bold lowercase and bold uppercase letters respectively. A vector function is denoted by a bold letter while a scalar function is denoted by a non-bold letter. Superscripts $(\cdot)^*$, $(\cdot)^{\intercal}$ and $(\cdot)^{H}$ stand for the conjugate, transpose and Hermitian transpose operators respectively. Subscripts $(\cdot)_m$ and $(\cdot)_k$ denote the antenna and user index. The expectation is denoted by $\mathbb{E}(.)$. The $M\times M$ identity matrix is given by $\mat{I}_M$. The main diagonal of a square matrix $\mat{A}$ is given by $\diag{\mat{A}}$. The trace of a matrix is given by $\mathrm{Tr}\left(\cdot\right)$. The element-wise or Hadamard product of two matrices is denoted by $\mat{A} \odot \mat{B}$. The element at location $(i,j)$ in matrix $\mat{A}$ is indicated as $[\mat{A}]_{i,j}$. $\mathcal{G} = (\mathcal{V}, \mathcal{E})$ denotes a graph where $\mathcal{V}$ is the set of nodes and $\mathcal{E}$ the set of edges. The edge going from node $a \in \mathcal{V}$ to node $b \in \mathcal{V}$ is denoted as $(a, b) \in \mathcal{E}$. The neighborhood of a node $a$ is denoted as $\mathcal{N}(a) = \{b \in \mathcal{V}: (a, b) \in \mathcal{E}\}$.

\section{Problem Formulation}
\label{sec:problem_forulation}

\begin{figure}
    \centering
    \includegraphics[width=0.75\textwidth]{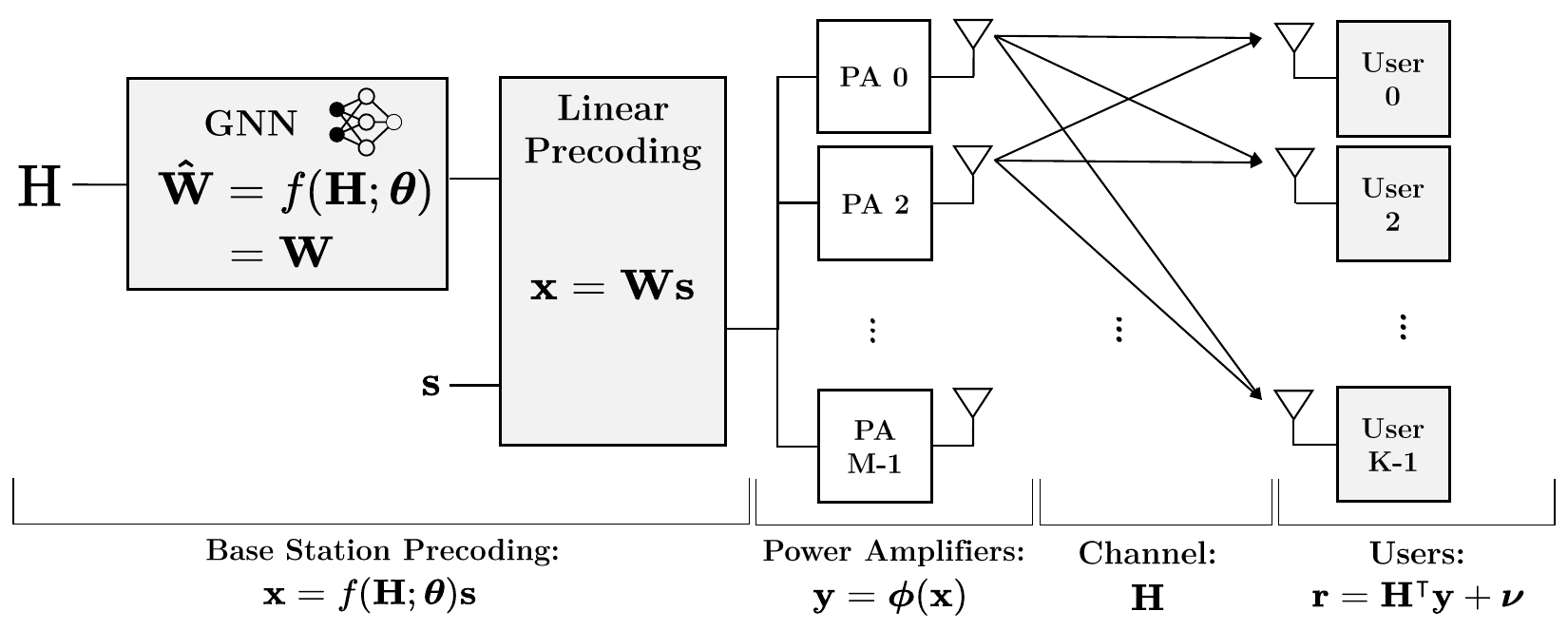}
    \caption{Massive MIMO downlink with neural network-based linear precoding and non-linear amplifiers at the transmitter.}
    \label{fig:overview}
\end{figure}
\subsection{System Model}
A downlink \gls{mmimo} system with non-linear \glspl{pa} and neural network-based linear precoding is considered as depicted in \cref{fig:overview}. In this system, $K$ single-antenna users are served by the \gls{bs} using $M$ transmit antennas. The complex symbol intended for user $k$ is denoted as $s_k$, it is assumed to be zero mean circularly symmetric complex Gaussian with unit variance. This is a valid assumption for e.g., \gls{ofdm} systems~\cite{ofdm}. The symbols for different users are assumed to be uncorrelated. The linearly precoded symbol at antenna $m$ is given by
\begin{align}
    x_m = \sum_{k=0}^{K-1} w_{m,k}s_k,
\end{align}
here $w_{m,k}$ is the precoding coefficient for user $k$ at antenna $m$. 
In matrix form, the precoded symbol vector $\vect{x} \in \mathbb{C}^{M\times1}$ is
\begin{align}
    \vect{x} = \mat{W}\vect{s}
\end{align}
where $\mat{W} \in \mathbb{C}^{M \times K}$ is the precoding matrix and $\vect{s} \in \mathbb{C}^{K \times 1}$ the symbol vector. The amplified transmit vector $\vect{y} \in \mathbb{C}^{M\times 1}$ is then given by 
\begin{align}
    \mat{y} = \boldsymbol{\phi} \left(\vect{x}\right)
\end{align} 
where $\boldsymbol{\phi}(\vect{x}) = \left[ \phi(x_0), \cdots, \phi(x_{M-1})\right]^{\intercal}$  denotes the element-wise non-linear transformation caused by the \glspl{pa}. Alternatively, in scalar notation the amplified signal at antenna $m$ can be denoted as $y_m = \phi(x_m)$.
The received signal vector $\vect{r} \in \mathbb{C}^{K\times1}$ is 
\begin{align*}
    \vect{r} &= \mat{H^\intercal} \vect{y} + \vect{v}
    =\mat{H^\intercal} \boldsymbol{\phi} \left(\mat{W} \vect{s}\right) + \vect{v}
\end{align*}
with $\mat{H} \in \mathbb{C}^{M \times K}$ being the channel matrix. The vector $\vect{v}\in \mathbb{C}^{K\times1}$ contains \gls{iid} zero mean complex Gaussian noise samples with variance $\sigma^2_v$.

\subsection{Non-Linear PA Models} \label{sec:pas}
The input signal of the \glspl{pa} is considered to be a bandpass signal, which is described by the following complex baseband representation
\begin{align}
   x(t) = A(t)e^{j \varphi(t)}
\end{align}
The output of a general non-linear PA with function $\phi(\cdot)$ is given by \cite{rf_imperfections_pas}
\begin{align}
   y(t) &= \phi(x(t)) \\
   &=\phi_{A}\left(A(t)\right) e^{j \varphi(t) + \phi_{\varphi}\left(A(t)\right)},
\end{align}
where $\phi_{A}(\cdot)$ and $\phi_{\varphi}(\cdot)$ represent the \gls{amam} and \gls{ampm} transfer function respectively.
In this work, the non-linear \gls{pa} is modeled as a complex valued polynomial of order $2N +1$. Furthermore, only the odd order polynomial coefficients are considered given that they contribute to spectral components near the carrier frequency, while the even order coefficients cause spectral components at multiples of the carrier frequency~\cite{rf_imperfections_pas}. Hence, the output of the \gls{pa} at antenna $m$ is given by
\begin{align}\label{eq:poly}
    \phi(x_m) &= \sum_{n=0}^N \beta_{2n+1} x_m |x_m|^{2n}  \\
    &= \beta_1 x_m + \beta_3 x_m |x_m|^2 + \cdots + \beta_{2N+1} x_m |x_m|^{2N}
\end{align}
where $\beta_{2n+1}$ are complex coefficients that model both \gls{amam} and \gls{ampm} distortion. Note that this formulation assumes that all $M$ \glspl{pa} have the same non-linear characteristic. 

In order to obtain proper polynomial coefficients at the desired \gls{ibo}, a least squares regression of the polynomial model to the modified Rapp model~\cite{modified_rapp} is performed. The \gls{amam} and \gls{ampm} distortion of the modified Rapp model are 
\begin{align}
    \phi_{A}(x_m) &= \frac{|x_m|}{\left(1+\left|\frac{x_m}{\sqrt{p_{\mathrm{sat}}}}\right|^{2S}\right)^{\frac{1}{2S}}} \label{eq:amamrapp}\\
    \phi_{\varphi}(x_m) &= \frac{A |x_m|^q}{1 + \left|\frac{x_m}{B}\right|^q}. \label{eq:ampmrapp}
\end{align}
The modified Rapp model coefficients are set as follows\footnote{Adapted from~\cite{3gpp} to a \gls{pa} with unit gain.}: $S=2$, $q=4$, $A=-0.315$, $B=1.137$ and the saturation power of the \gls{pa}, $p_{\mathrm{sat}}$ is scaled in order to produce the desired \gls{ibo} according to $\mathrm{IBO} = p_{\mathrm{in}}/p_{\mathrm{sat}}$, with $p_{\mathrm{in}}$ being the average input power at each \gls{pa}. 

\begin{figure}
    \centering
    \includegraphics{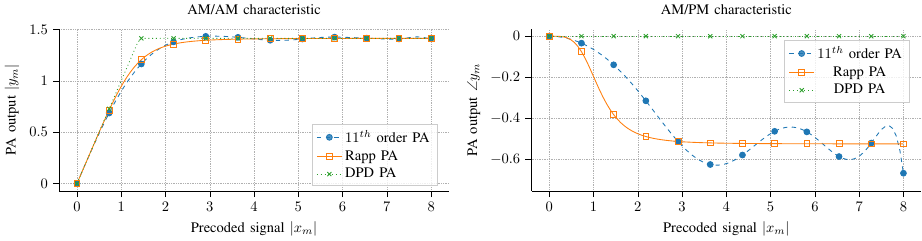}
    \caption{\gls{amam} and \gls{ampm} characteristics of the considered power amplifier models: modified Rapp model, $11^{\mathrm{th}}$ order polynomial model and a soft limiter/perfect \gls{dpd}.}
    \label{fig:pas}
\end{figure}

Additionally, as a benchmark a perfect \gls{dpd} is considered. This is modeled as a soft limiter, i.e., a linear \gls{amam} characteristic up to a certain saturation point where the output of the \gls{pa} is clipped~\cite{rf_imperfections_pas}, which is equivalent to the \gls{amam} characteristic of the Rapp model for $S$ going to infinity. The \gls{amam} characteristic is thus modeled as
\begin{align}\label{eq:dpd}
     \phi_{A}(x_m) &= \begin{cases}|x_m| & \text { for } \quad |x_m| \leq \sqrt{p_{\mathrm{sat}}} \\ \sqrt{p_{\mathrm{sat}}} & \text { for } \quad |x_m|>\sqrt{p_{\mathrm{sat}}}\end{cases},
\end{align}
while the \gls{ampm} conversion is zero. An overview of these \glspl{pa} models is given in \cref{fig:pas}. 

\subsection{Radiation Patterns}
In \cref{sec:results}, at some places, the radiation pattern of the proposed precoder is evaluated. For this, a pure \gls{los} channel is considered. The channel from antenna $m$ to user $k$ can then be written as
\begin{align}
    h_{m,k} = \sqrt{\beta_k} e^{-j m \frac{2 \pi}{\lambda_c} d \cos(\theta_k)}.\label{eq:los}
\end{align}
Here, $\beta$ models the path loss, while $m \frac{2 \pi}{\lambda_c} d \cos(\theta_k)$ represents the antenna dependent phase shift when considering a \gls{ula} and a narrowband system. Additionally, $\lambda_c$ denotes the carrier wavelength, $\theta_k$ the user angle and $d$ the antenna spacing. The radiation pattern in an arbitrary direction $\Tilde{\theta}$ can be obtained by
\begin{align}
    P(\Tilde{\theta}) = \expt{\left| \sum_{m=0}^{M-1} y_m e^{-j m \frac{2 \pi}{\lambda_c} d \cos(\Tilde{\theta})}\right|^2}. \label{eq:rad}
\end{align} 
The radiation pattern of the linearly amplified signal is obtained by replacing $y_m$ by $x_m$ in (\ref{eq:rad}) and is denoted as $P_{\mathrm{lin}}(\Tilde{\theta})$. The radiation pattern of the non-linear distortion is denoted as $P_{\mathrm{dist}}(\Tilde{\theta})$ and is obtained by replacing $y_m$ by $\sum_{n=1}^N \beta_{2n+1} x_m |x_m|^{2n}$ in (\ref{eq:rad}). Additionally, the \gls{sdr} radiation pattern is defined as $P_{\mathrm{SDR}}(\Tilde{\theta}) = \frac{P_{\mathrm{lin}}(\Tilde{\theta})}{P_{\mathrm{dist}}(\Tilde{\theta})}$. %

\subsection{Problem Formulation}
An achievable sum rate $R_{\mathrm{sum}}$, i.e., a lower bound on the capacity, is obtained by considering that the noise and distortion are jointly Gaussian distributed and independent from the data symbols, this can be seen as a worst case
\begin{align}\label{eq:rate}
        R_{\mathrm{sum}} = \sum_{k=0}^{K-1} \mathrm{log}_2(1 + \mathrm{SNIDR}_k),
\end{align}
where $\mathrm{SNIDR}_k$ is the \gls{snidr} at user $k$. It can be computed based on the Bussgang decomposition~\cite{demir2020bussgang}, which implies that the received signal for user $k$ can be written as $  r_{k} = {B_{k}  s_k} + d_{k} + v_{k}$. Here, $d_k$ captures both the non-linear distortion and inter-user interference, which is uncorrelated to the transmit signal $s_k$ and the noise $v_k$. The linear gain is given by $B_{k}=\mathbb{E}\left(r_{k} s_k^{*}\right) / p_k$, with $p_k = \mathbb{E}(s_ks_k^*)$~\cite{demir2020bussgang}. The received signal variance for user $k$ is given by $|B_k|^2p_k$. The distortion and inter-user interference can be computed as $\mathbb{E}\left(|d_{k} |^{2}\right)=\mathbb{E}\left(|r_{k} |^{2}\right)-|B_{k} |^{2} p_k-\sigma_{v_{k}}^{2}$, given that $d_k$, $s_k$ and $v_k$ are uncorrelated. The \gls{snidr} for user $k$ is then given by
\begin{align}
    \mathrm{SNIDR}_k = \frac{|B_k |^{2} p_k}{\mathbb{E}\left(|d_{k}|^{2}\right)+\sigma_{v}^{2}}. \label{eq:snidr_num}
\end{align}
This general expression for the \gls{snidr} can be evaluated numerically for a general \gls{pa} model and is used for the evaluation in~\cref{sec:results}. 

For training the \glspl{nn}, the $(2N+1)^{\mathrm{th}}$ order polynomial \gls{pa} model in (\ref{eq:poly}) is assumed, which leads to an analytical expression for the \gls{snidr}~\cite{dab_mmwave, dist_derivation}. By applying Bussgang's theorem\cite{demir2020bussgang} to the amplification stage, we can write the amplified signal as
\begin{align}
    \boldsymbol{\phi}(\vect{x}) = \mat{G} \vect{x} + \vect{e},
\end{align}
with $\vect{e}\in \mathbb{C}^{M\times1}$ the non-linear distortion term and $\mat{G}\in \mathbb{C}^{M\times M}$ a diagonal matrix containing the Bussgang gains with the diagonal entries being $[\mat{G}]_{m,m} = \mathbb{E}[\phi(x_m)x_m^*]/\mathbb{E}[|x_m|^2]$.

When assuming the $(2N+1)^{\mathrm{th}}$ order polynomial model and linear precoding ($\vect{x} = \mat{W}\vect{s}$), the gain matrix $\mat{G}$ can be written as a function of the precoding matrix $\mat{W}$~\cite{dist_derivation}
\begin{align}
    \mat{G}(\mat{W}) &= \sum_{n=0}^N (n+1)! \beta_{2n+1} \mat{I}_M \diag{\mat{C}_x}^n \label{eq:gainmat}
\end{align}
The input covariance matrix is given by $\mat{C}_x = \mathbb{E}[\vect{x}\vect{x}^H] = \mat{W}\mat{W}^H$. From~\cite{dab_mmwave, dist_derivation}, the covariance matrix of the non-linear distortion $\vect{e}$ can be derived as
\begin{align}
    \mat{C}_{\vect{e}}(\mat{W}) = \sum_{n=1}^N \mat{L}_n \mat{C}_x \odot |\mat{C}_x|^{2n} \mat{L}_n^{H},\label{eq:cov_e}
\end{align}
with
\begin{align}
    \mat{L}_n = \frac{1}{\sqrt{n+1}} \sum_{l=n}^N \binom{l}{n} (l+1)! \beta_{2l+1} \mat{I}_M \diag{\mat{C}_x}^{l-n}. \label{eq:ln}
\end{align}
The received signal at user $k$ can then be written as
\begin{align}
    r_k &= \underbrace{\vect{h}_k^\intercal \mat{G(W)} \vect{w}_k s_k}_{\text{desired signal}} + \underbrace{\sum_{k'\neq k} \vect{h}_k^{\intercal} \mat{G(W)}\vect{w}_{k'}s_{k'}}_{\text{inter-user interference}} + \underbrace{\vect{h}_k^{\intercal} \vect{e}}_{\substack{\text{received} \\ \text{non-linear distortion}}} + \underbrace{v_k}_{\text{noise}}.
\end{align}
This leads to the following \gls{snidr} expression for user $k$
\begin{align}\label{eq:snidr_analytical}
    &\mathrm{SNIDR}_k(\mat{W}) = \frac{|\vect{h}_k^{\intercal} \mat{G(W)} \vect{w}_k|^2}{\sum\limits_{k'\neq k}|\vect{h}_k^{\intercal} \mat{G(W)} \vect{w}_{k'}|^2 + \vect{h}_k^{\intercal} \mat{C}_e(\mat{W}) \vect{h}^*_k + \sigma_v^2}.
\end{align}
Given this expression for the \gls{snidr}, an achievable sum rate can be computed using (\ref{eq:rate}).
As such, the optimization problem we aim to solve can be formulated as
\begin{align}\label{eq:opt}
    \max_{\mat{W}} \quad & R_{\mathrm{sum}}\left(\mat{W}\right) \\
    \textrm{s.t.} \quad & \mathbb{E}\left(\sum_{m=0}^{M-1}|x_m|^2\right) =  \mathrm{Tr}\left(\mat{W}\mat{W}^H\right) \leq P_T, \nonumber
\end{align}
where $P_T$ is the total transmit power. The aim is thus to find a precoding matrix which maximizes the sum rate, subject to a power constraint\footnote{For simplicity, the power constraint is taken before the \gls{pa}, which neglects the non-linearly amplified power, which is small compared to the full transmit power. Due to the saturation characteristic of the PA, this can be seen as an upper bound on the actual transmit power.}, while the system is affected by PA non-linearities.

\subsection{Neural Network Training}
The neural network-based precoder considered in this work takes as input the channel matrix $\mat{H}$ and outputs the linear precoding matrix $\mat{\hat{W}}$. This neural network can be represented as $\mat{\hat{W}} = f(\mat{H}; \mat{\theta})$ i.e., a learned non-linear function mapping which is parameterized by $\mat{\theta}$. Training is done in a self-supervised manner by maximizing the sum rate in order to obtain the neural network parameters
\begin{align}
    \mat{\theta}^* = \argmin_{\mat{\theta}} - R_{\mathrm{sum}}(f\left(\mat{H}; \mat{\theta}\right)).
\end{align}
Here the sum rate $R_{\mathrm{sum}}$ is given by (\ref{eq:rate}), where the \gls{snidr} is computed according to (\ref{eq:snidr_analytical}). Since the \gls{snidr} in (\ref{eq:snidr_analytical}) directly depends on the output of the neural network (i.e., the precoding matrix $\mat{\hat{W}}$), the gradients of this loss function, with respect to the neural network parameters $\mat{\theta}$, can directly be computed using backpropagation. Given these gradients, the parameters of the \gls{nn} are updated using the Adam optimizer\cite{adam}. %

\subsection{Benchmark Algorithms}
Throughout this study, the proposed solution is compared against a number of benchmark algorithms. 
First, \gls{mrt} and \gls{zf} precoding are considered~\cite{fundamentals_mimo}
\begin{align}
    \mat{W}^{\mathrm{MRT}} &= \alpha \mat{H}^*\\
    \mat{W}^{\mathrm{ZF}} &= \alpha \mat{H}^* (\mat{H}^{\intercal} \mat{H}^*)^{-1},\label{eq:zf}
\end{align}
where $\alpha = \sqrt{P_T/\mathrm{Tr}(\mat{W} \mat{W}^H)}$ is a power normalization constant. The \gls{mrt} and \gls{zf} precoding matrices are obtained under the assumption that the \gls{pa} is linear. Hence, these two algorithms act as a benchmark in the linear regime where no distortion is present. Additionally, they represent a worst-case scenario, highlighting how the system's performance deteriorates when non-linear distortion is present and no mitigation techniques are adopted. 

Second, the \gls{z3ro} precoder from~\cite{z3ro} is considered. This precoder was designed for the single-user case, assuming a third-order \gls{pa} model. It maximizes the \gls{snr} while nulling the third-order distortion term at the user location. This is done by saturating a number of antennas with an opposite phase shift, as can be seen in (\ref{eq:z3ro}). The closed-form expression for the precoding coefficient at antenna $m$ is given by
\begin{align}
    w_m^{\mathrm{Z3RO}, M_s}=\alpha h_m^* \begin{cases}-\gamma & \text { if } m=0, \ldots, M_s-1 \\ 1 & \text { otherwise }\end{cases}, \label{eq:z3ro}
\end{align}
here $\gamma=\left(\nicefrac{\sum_{m^{\prime}=M_s}^{M-1}\left|h_{m^{\prime}}\right|^4}{\sum_{m^{\prime \prime}=0}^{M_s-1}\left|h_{m^{\prime \prime}}\right|^4}\right)^{1 / 3}$ is the gain of the saturated antennas and $M_s$ is the number of saturated antennas, which is set to $M_s = 1$ in all simulations, as this gives the globally optimal solution~\cite{zerofamily}. In~\cite{zerofamily}, this precoder was shown to be optimal in the single-user case and when a third-order \gls{pa} model is assumed. Hence, in this study, it is used to evaluate how close the \gls{nn} precoder is to the optimal solution, under these assumptions. 

For the more general multi-user case and when high-order \glspl{pa} are assumed, the problem becomes non-convex, consequently, no optimal solution is known. However, in~\cite{dab_mmwave} the non-convex optimization problem is approximately solved using an iterative projected gradient ascent method, which is denoted as the \gls{dab} precoder. For the multi-user and high-order \gls{pa} case, this algorithm is used as a benchmark. The precoding matrix obtained by this algorithm is denoted as $\mat{W}^{\mathrm{DAB}}$ and can be obtained by performing the following algorithm for $I$ iterations
\begin{align}
    \mat{W}^{(i+1)} = \left[ \mat{W}^{(i)} + \mu^i \nabla_{\mat{W}} R_{\mathrm{sum}}\left(\mat{W}^{(i)}\right)\right]_{\mathbb{E}\|\boldsymbol{\phi}(\mat{W}\vect{s})\|^2=P_T}^+ \label{eq:dab}
\end{align}
Here $\mu^i$ is the step size for the gradient update at iteration $i$ and $[\cdot]_{\mathbb{E}\|\boldsymbol{\phi}(\mat{W}\vect{s})\|^2=P_T}^+$ denotes the projection of the obtained solution, such that the power constraint is met\footnote{Note that in~\cite{dab_mmwave} the power constraint is considered after the \gls{pa} in contrast to our approach where the power constraint is enforced before the \gls{pa}. This implies that our proposed precoder abides by both power constraints, i.e., before and after the \gls{pa}, while the \gls{dab} precoder only abides by the constraint after the \gls{pa}.}.

\section{Neural Networks for Precoding}
\label{sec:nn_for_prec}

\subsection{Suitable Neural Network Architectures}\label{sec:nn_comp}

Neural networks can approximate any continuous non-linear function with arbitrary accuracy~\cite{universalapprox}. Subsequently, they can be used to learn a mapping from channel matrix $\mat{H}$ to precoding matrix $\mat{\hat{W}}$. 
Unfortunately, the universal approximation theorem is not constructive, i.e., it does not provide a way to select the \gls{nn} architecture that can achieve this arbitrary accuracy. A \gls{mlp} is a general fully connected feedforward neural network, which is a universal function approximator~\cite{universalapprox}. As such, in theory, \glspl{mlp} could be used to learn the mapping between channel and precoding matrix. In practice, because of the general structure of \glspl{mlp}, this leads to a very large number of trainable parameters. Consequently, these \glspl{mlp} are very hard to train and have a high computational complexity at inference time. As such, it is advantageous to select a \gls{nn} architecture, with a certain inductive bias, that suits the learning task. The inductive bias limits the types of functions that can be learned, thus decreasing the size of the hypothesis space that the \gls{nn} covers. If this inductive bias corresponds to the learning task, the desired function is still covered by the (reduced) hypothesis space of the \gls{nn}. This produces more scalable \glspl{nn}, with fewer learnable parameters, that are easier to train and have a lower computational complexity~\cite{inductivebiasold}.

In order to reduce the hypothesis space for the precoding problem the permutation equivariance property of precoding can be utilized. From~\cite{cnn_gnn}, it is clear that precoding is permutation equivariant with respect to the users and antennas. To clarify, if the order of the users or antennas in the channel matrix $\mat{H}$ is permuted, the order of the precoding vectors in $\mat{W}$ should be permuted accordingly, but the sum rate remains unchanged. More formally, a precoding function $\mat{W} = f(\mat{H})$ is permutation equivariant if the following holds $ f(\mat{\Pi}_1 \mat{H} \mat{\Pi}_2^{\intercal}) = \mat{\Pi}_1 f(\mat{H}) \mat{\Pi}_2^{\intercal} $, where $\mat{\Pi}_1,  \mat{\Pi}_2$ are permutation matrices~\cite{cnn_gnn}. Hence, when a \gls{nn} architecture naturally exhibits this permutation equivariance property, it is beneficial for the precoding task. 
\Glspl{ccnn} are shift equivariant~\cite{cnn_gnn}, more formally this implies that $f_{\mathrm{CCNN}}(\mat{S}_1 \mat{H} \mat{S}_2^{\intercal}) = \mat{S}_1 f_{\mathrm{CCNN}}(\mat{H}) \mat{S}_2^{\intercal} $, where $\mat{S}_1$ and $\mat{S}_2$ are shift matrices i.e., circulant permutation matrices~\cite{cnn_gnn}. In~\cite{cnn_gnn} it is shown that \glspl{ccnn} can learn permutation equivariant functions for certain values of the learned weight matrices. However, in general, the functions learned by the \gls{ccnn} are not permutation equivariant. This implies that the inductive bias of the \gls{ccnn} is more general than solely consisting of permutation equivariant functions. This means that the hypothesis space of a \gls{ccnn} is larger than that of a \gls{nn} that learns purely permutation equivariant functions. In our previous work~\cite{icc_ccnn}, we showed that these \glspl{ccnn} can be used to learn precoding functions that can cancel third-order non-linear distortion. In the current study, we show that \glspl{gnn} can produce even better results, in the presence of both third and higher order non-linear distortion. \glspl{gnn} are naturally permutation equivariant~\cite{graph_rep_learning}, which gives them the appropriate inductive bias for precoding.

\begin{figure}
    \centering
    \subfloat[]{\label{fig:mlp_cnn_gnn}\includegraphics[]{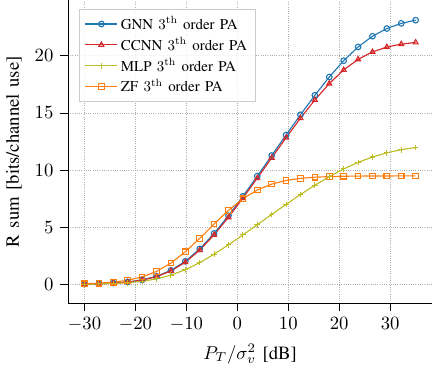}}
    \hfill
    \subfloat[]{\label{fig:compare_learnable_params}\includegraphics{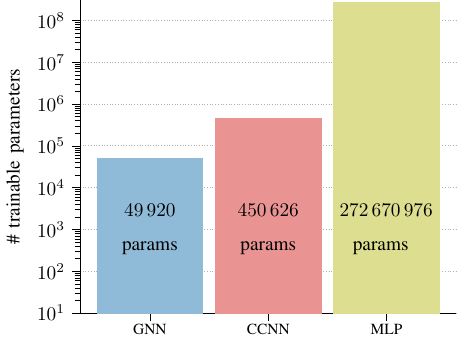}}
    \caption{(a) Comparison of an \gls{mlp}, \gls{ccnn} from~\cite{icc_ccnn} and \gls{gnn} for $M=64$, $K=2$. Trained and evaluated on a third-order non-linear \gls{pa} at $-3$ dB \gls{ibo}, with $\beta_3 = -0.07781605 - 0.0401193j$. (200K training samples, 10K test samples) (b) Comparison of the number of trainable parameters of these three architectures.}
    \label{fig:mlp_cnn_gnn_comp}
\end{figure}

To illustrate this,~\cref{fig:mlp_cnn_gnn} denotes the performance of the different \gls{nn} architectures for a scenario where $M=64$, $K=2$ and third-order non-linear amplifiers are considered. All 3 architectures are matched in terms of their representation capacity i.e., each of them has 6 layers and 64 features\footnote{64 features indicates that the real and imaginary part of each precoding coefficient is represented by a feature vector of dimension 64, resulting in a total of $64MK2$ features.} per layer. From this image it is clear that the \gls{gnn} achieves the highest sum rate as compared to the \gls{ccnn} and \gls{mlp}. The \gls{ccnn} achieves a slightly lower sum rate, because of the shift invariance it captures, which is still a relatively good inductive bias for precoding. The \gls{mlp} has a very general structure and as such can barely produce a sum rate that is higher than that of the \gls{zf} precoder. Moreover, in \cref{fig:compare_learnable_params} the number of learnable parameters of each \gls{nn} architecture is illustrated. This again highlights the proficiency of the \gls{gnn} as it only has \num{50000} learnable parameters as compared to the \num{450000} and \num{270000000} of the \gls{ccnn} and \gls{mlp}. 

\subsection{Graph Neural Network Precoding}\label{sec:gnn_basic}
\Glspl{gnn} generally come in two forms, namely \glspl{gnn} that learn a hidden representation of the nodes in a graph and \glspl{gnn} that learn a hidden representation of the edges in a graph. For the precoding problem, we focus on \glspl{gnn} that learn representations of the edges of the graph. In general, the goal is thus to start from a number of input edge features and learn edge embeddings. Such a \gls{nn} consists of a number of layers or message-passing iterations, where each layer/iteration produces a new representation/embedding of the edges. An embedding/representation of edge $(a,b)$ in layer $l$ of a \gls{gnn} is expressed as $\vect{z}_{(a,b)}^{(l)}$. A typical \gls{gnn} layer consists of two steps, namely, a message-passing/aggregation step and an update step. Hence, a single layer for edge $(a,b)$ can be expressed as 
\begin{align}
    \vect{z}_{(a,b)}^{(l+1)} = \mathrm{UPDATE}^{(l)}\big( &\vect{z}_{(a,b)}^{(l)} ,\vect{m}^{(l)}_{\mathcal{N}(a)}, \vect{m}^{(l)}_{\mathcal{N}(b)}
    \big)
\end{align}
Here $\vect{m}_{\mathcal{N}(a)}, \vect{m}_{\mathcal{N}(b)}$ are 'messages' that are aggregated from the neighboring edges of the two nodes connected to edge $(a,b)$, these messages are defined as
\begin{align}
    \vect{m}^{(l)}_{\mathcal{N}(a)} &=  \mathrm{AGGREGATE}^{(l)}\left(\{\vect{z}_{(a,x)}^{(l)}, \forall x \in \mathcal{N}(a)\}\right)\\
    \vect{m}^{(l)}_{\mathcal{N}(b)}& = \mathrm{AGGREGATE}^{(l)}\left(\{\vect{z}_{(b,x)}^{(l)}, \forall x \in \mathcal{N}(b)\}\right).
\end{align}
Here '$\mathrm{UPDATE}$' and '$\mathrm{AGGREGATE}$' are differentiable functions. For a comprehensive overview of commonly used update and aggregate functions we refer to~\cite{graph_rep_learning}.

\begin{figure}[t]
    \centering
    \begingroup
    \tikzset{every picture/.style={scale=0.9}}%
    \subfloat[Message passing from users to antennas in layer $l$ for edge $(0, 0)$.]{\label{fig:overview_a}\includegraphics{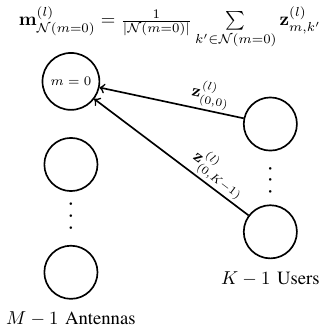}}\hfill
    \subfloat[Message passing from antennas to user in layer $l$ for edge $(0, 0)$.]{\label{fig:overview_b}\includegraphics{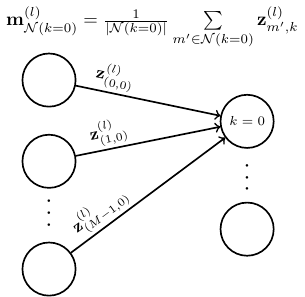}}\hfill
     \subfloat[Update the edge feature using learned weight matrices for edge $(0, 0)$.]{\label{fig:overview_c}\includegraphics{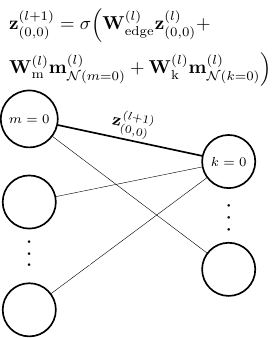}}
     \endgroup
	\caption{\Gls{gnn} visualized for a single edge $(0, 0)$ in the graph. In (a) and (b) message passing is performed. Note that in the first layer $l=0$ the channel coefficients act as input edge features i.e., $ \mathbf{z}_{(m, k)}^{(0)} = \begin{bmatrix} \Re \{h_{m,k}\} , \Im \{h_{m,k} \}\end{bmatrix}^{\intercal}$. In (c) the messages and edge features are used to update the edge feature which acts as the output of this layer. This is repeated for each layer to generate the outputs of the network. The outputs of the network are the edge features in the final layer which represent the precoding coefficients $ \mathbf{z}_{(m, k)}^{(L-1)} = \begin{bmatrix} \Re \{w_{m,k}\} , \Im \{w_{m,k} \}\end{bmatrix}^{\intercal} $ .  (Note that the power normalization layer is omitted for simplicity.)}%
	\label{fig:overview_gnn} 
\end{figure}

More specifically, the precoding problem can be represented as learning a mapping from a $M \times K$ MIMO channel matrix to a $M\times K$ precoding matrix over a graph, a visualization of this is given in~\cref{fig:overview_gnn}. The graph can be represented as follows: each antenna $m$ and user $k$ is denoted by a node. The edges between the antennas and users represent the wireless channel and have as input features the channel coefficients. This graph is defined as $\mathcal{G} = (\mathcal{V}, \mathcal{E})$, where $\mathcal{V}$ is the set of nodes and $\mathcal{E}$ the set of edges. $\mathcal{V
}$ contains all user and antenna nodes, i.e., $m \in \mathcal{V} \ \forall m$ and $k \in \mathcal{V} \ \forall k$. $\mathcal{E}$ contains all edges between the antennas and users i.e., $(m, k) \in \mathcal{E} \ \forall m, k \in \mathcal{V}$. This graph is undirected, meaning that $(m, k) \in \mathcal{E} \leftrightarrow (k, m) \in \mathcal{E}$, as is the physical channel between each user and \gls{bs} antenna. For this graph, a single \gls{gnn} layer that updates the representation of edge $(m,k)$ can be defined as
\begin{align}
    \vect{z}^{(l+1)}_{(m,k)} &= \mathrm{UPDATE}_{\mathrm{edge}}\left(\vect{z}^{(l)}_{(m,k)}, \vect{m}^{(l)}_{\mathcal{N}(m)}, \vect{m}^{(l)}_{\mathcal{N}(k)}\right) \\
    &= \sigma \left(\mat{W}^{(l)}_{\mathrm{edge}} \vect{z}^{(l)}_{(m,k)} + \mat{W}^{(l)}_{m}\vect{m}^{(l)}_{\mathcal{N}(m)} + \mat{W}^{(l)}_{k}\vect{m}^{(l)}_{\mathcal{N}(k)} \right), \label{eq:updatebase}
\end{align}
where $\vect{z}^{(l+1)}_{(m,k)}$ denotes the hidden representation of the edge $(m,k)$ at layer/iteration  $l+1$ of the \gls{gnn}, $\sigma(\cdot)$ is a non-linear activation function and the matrices $\mat{W}^{(l)}_{\mathrm{edge}}, \mat{W}^{(l)}_m, \mat{W}^{(l)}_k \in \mathbb{R}^{d_{l+1} \times d_{l}}$ are learnt using a form of stochastic gradient descent. Note that the dimensions of these matrices, $d_{l+1}$ and $d_{l}$, determine the number of features in layer $l+1$ and $l$ respectively. The aggregation operations are illustrated in \cref{fig:overview_a} and \ref{fig:overview_b} and are defined as follows
\begin{align}
    \vect{m}^{(l)}_{\mathcal{N}(m)} &= \frac{1}{\left|\mathcal{N}(m)\right|}\sum_{k'\in\mathcal{N}(m)} \vect{z}^{(l)}_{(m,k')} \label{eq:messagem}\\ 
    \vect{m}^{(l)}_{\mathcal{N}(k)}& = \frac{1}{\left|\mathcal{N}(k)\right|}\sum_{m'\in\mathcal{N}(k)} \vect{z}^{(l)}_{(m',k)}. \label{eq:messagek}
\end{align}
The input and output for edge $(m,k)$ of the network are defined in the following way 
\begin{align}
    \mathbf{z}_{(m, k)}^{(0)} &= \begin{bmatrix} \Re \{[\mathbf{H}]_{m,k}\} , \Im \{[\mathbf{H}]_{m,k} \}\end{bmatrix}^{\intercal}\\
    \mathbf{z}_{(m, k)}^{(L-1)} &= \begin{bmatrix} \Re \{[\mathbf{\hat{W}}]_{m,k}\} , \Im \{[\mathbf{\hat{W}}]_{m,k} \}\end{bmatrix}^{\intercal}.
\end{align}
In order to respect the power constraint, a power normalization layer is added, which consists of a scalar normalization given by 
\begin{align}
    \mat{\hat{W}}^{norm} = \alpha \mat{\hat{W}}, \label{eq:pwr_norm}
\end{align}
with $\alpha = \sqrt{P_T/\mathrm{Tr}(\mat{\hat{W}} \mat{\hat{W}}^H)}$.

\subsection{Leveraging SNR Regime Information}\label{sec:snrgnn}
For the neural network architecture as described in the~\cref{sec:gnn_basic}, the network is typically trained at a fixed \gls{snr} ($=P_{T}/\sigma_{\nu}^2$) point. This results in the fact that the performance of the neural network is best close to the \gls{snr} value it is trained at. In this section, it is outlined how the \gls{snr} can be included as an input edge feature. This allows the neural network to be trained at a number of \gls{snr} points, depending on the input the neural network can differentiate in which SNR regime it is operating and deliver the best performance. To do so the input to the \gls{gnn} is defined in the following way 
\begin{align}
     \mathbf{z}_{(m, k)}^{(0)} &= \begin{bmatrix} \Re \{[\mathbf{H}]_{m,k}\} , \ \Im \{[\mathbf{H}]_{m,k}\}, \  \mathrm{SNR}_{\mathrm{norm}}  \}\end{bmatrix}^{\intercal}.
\end{align}
To ensure stability during training we include the normalized \gls{snr} as input rather than the absolute \gls{snr}. The normalized SNR is defined as follows
\begin{align}
    \mathrm{SNR}_{\mathrm{norm}} = \frac{1}{\mathrm{SNR}_{\mathrm{max}}} \frac{P_t}{\sigma^2_{\nu}}.
\end{align}
Training is done for an SNR ($=P_T/\sigma_{\nu}^2$) range of -30 to 30 dB, with a stepsize of 5 dB, hence $\mathrm{SNR}_{\mathrm{max}} = 30$ dB. During training, the SNR for each training example is randomly selected from this range. During testing, the sum rate is evaluated for an SNR range of -30 to 30 dB, however this time with a spacing of 2.5 dB. This ensures that SNR points for which the model is not trained are also tested. %

\section{Complexity and Real-Time Analysis}
\label{sec:complexity}

In this section, the complexity of the proposed solution is quantified and compared against the benchmark precoders. Next to this, it is outlined which operations of the \gls{gnn} can be performed in parallel in order to produce faster execution times for real-time implementations. In order to quantify the computational complexity of the proposed solution, the number of real floating point additions and multiplications are computed. Note that an addition and multiplication carry a different computational cost with the multiplication being the most costly one. However this highly depends on the hardware platform~\cite{hardware_survey}. As such, both the number of real floating point additions and multiplications are computed. Additionally the number of \gls{flops} is computed. A FLOP is defined as either a real floating point multiplication or addition. When considering complex multiplications or additions a conversion to real multiplications and additions is made. A complex multiplication requires 4 real multiplications and 2 real additions, while a complex addition requires 2 real additions. Note that the number of computations needed to perform the power normalization of the precoders is neglected, as this is equal across the precoders\footnote{Note that the \gls{dab} precoder employs a more complex power normalization which is performed after the amplification stage. However, a more simple power normalization before the \gls{pa} could be used for this precoder, which would render the complexity of the normalization across all precodes equal. As such we decided not to include this term into the complexity analysis.}. An overview of the complexity analysis can be found in~\cref{tab:complexity}.

\subsection{Complexity of the GNN}
In order to perform one forward pass of the \gls{gnn} (\ref{eq:updatebase}) has to be computed $L$ times, namely the number of layers in the \gls{gnn}, for each edge $(m, k) \in \mathcal{E}$ i.e., $MK$ times. Note that our \gls{gnn} architecture has $L=8$ layers, meaning it consists out of an input layer, 6 hidden layers and 1 output layer. The number of in- and output features is fixed to $d_0 = d_{L-1} = 2$ and the number of features in each hidden layer is constant at $d_l = 128 \; \forall l \in \{2, \cdots, L-2\}$.
The \gls{gnn} first performs message passing as illustrated in \cref{fig:overview_a} and \ref{fig:overview_b} i.e., computing $\vect{m}^{(l)}_{\mathcal{N}(m)}$ and $\vect{m}^{(l)}_{\mathcal{N}(k)}$, which can be done by computing (\ref{eq:messagem}) and (\ref{eq:messagek}) this has the following complexity:
\begin{itemize}
    \item Input layer: $2 (M-1) + 2 (K-1)$ real floating point additions
    \item Output and Hidden layers: $(L-1) (d_l (M-1) + d_l (K-1))$ real floating point additions
\end{itemize}
Subsequently, the update step is performed as illustrated in~\cref{fig:overview_c}. This involves a number of matrix multiplications with the learned weight matrices, followed by the addition of these products, this has the following complexity:
\begin{itemize}
    \item Input and output layer: $12 d_l$  real multiplications and $11 d_l - 2$  real additions 
    \item Hidden layers: $(L-2) 3d_l^2$  real multiplications and $(L-2) (3d_l^2-d_l) $  real additions 
\end{itemize}
Note that the complexity of the non-linear activation function is neglected as it is often approximated based on lookup tables, which requires few computations~\cite{lut_activations}.

In total the \gls{gnn} needs the following number of real floating point multiplications
\begin{align}
    \underbrace{MK}_{\text{edges}} \bigg( \underbrace{\overbrace{6d_l}^{\text{(a)}}}_{\text{input layer}} + \underbrace{(L-2)\overbrace{3d_l^2}^{\text{(a)}}}_{\text{hidden layers}} + \underbrace{\overbrace{6d_l}^{\text{(a)}}}_{\text{ouput layer}}\bigg)
\end{align}
and the following number of real floating point additions
\begin{align}
    \underbrace{MK}_{\text{edges}} \bigg( \underbrace{\overbrace{5 d_l}^{\text{(a)}} + \overbrace{2(M-1+K-1)}^{\text{(b)}}}_{\text{input layer}} + \underbrace{(L-2)\Big(\overbrace{3d_l^2 -d_l}^{\text{(a)}} + \overbrace{d_l(M-1+K-1)}^{\text{(b)}}\Big)}_{\text{hidden layers}} + \underbrace{\overbrace{6d_l-2)}^{\text{(a)}}}_{\text{output layer}}\bigg).
\end{align}
where the term (a) denotes the computations needed to perform the matrix multiplication with the learned weight matrices followed by the addition of the resulting products and (b) denotes the computations needed for the message passing. 
This scales as $\mathcal{O}\left(MKd_l^2L \right)$ floating point multiplications and $\mathcal{O}\left(MKd_l^2L + M^2Kd_lL + MK^2d_lL \right) $ floating point additions. In terms of \gls{flops} this corresponds to $\mathcal{O}\left(MKd_l^2L + M^2Kd_lL + MK^2d_lL \right)$. 
Ultimately, it is important to highlight that the utilization of \glspl{nn} offers a distinct advantage. Namely, a substantial portion of the computational tasks can be effectively parallelized, thereby resulting in a faster execution time. In our specific scenario, the execution of (\ref{eq:updatebase}) can be parallelized across all edges, yielding a reduced number of \gls{flops} that need to be executed serially, namely $\mathcal{O}\left(d_l^2L + Md_lL + Kd_lL \right)$. This can be leveraged to achieve real-time operation.

\subsection{Complexity of Zero Forcing Precoder}
In order to compute the \gls{zf} precoder as given in (\ref{eq:zf}), first a matrix multiplication of a $K \times M$ with an $M \times K$ matrix is computed, which requires $MK^2$ complex multiplications and $K^2(M-1)$ complex additions, this corresponds to $2MK^2-K^2$ \gls{flops}. Second, the inverse of a $K \times K$ matrix needs to be computed, which can efficiently be done by first computing a Cholesky factorization, followed by taking the inverse of the resulting lower triangular matrix~\cite{flops2}. This requires $\frac{1}{2}K^3 + \frac{3}{2}K^2$ complex multiplications, $\frac{1}{2}K^3-\frac{1}{2}K^2$ complex additions and the computation of $K$ square-root operations, which is equivalent to $K^3 + K^2 + K$ complex \gls{flops}. Finally, a matrix multiplication of an $M \times K$ with an $K \times K$ matrix needs to be computed, which requires $MK^2$ complex multiplications and $MK(K-1)$ complex additions which results in $2MK^2 - MK$ complex \gls{flops}. This results in a total number of $2MK^2 + \frac{1}{2}K^3 + \frac{3}{2}K^2$ complex multiplications and $2MK^2 + \frac{1}{2}K^3-\frac{3}{2}K^2$ complex additions. This is equivalent to $8MK^2 + 2K^3 + 6K^2$ real multiplications, $8MK^2 + 2K^3$ real additions or $16MK^2+4K^3+6K^2$ real \gls{flops}. This scales as $\mathcal{O}\left(MK^2 + K^3\right)$ \gls{flops}.

\subsection{Complexity of Distortion-aware Beamforming Precoder}
The \gls{dab} precoder executes an iterative optimization procedure for $P=50$ different initializations. For each initialization, (\ref{eq:dab}) is computed for $I=1000$ iterations. This requires $2MK$ real additions and $2MK$ real multiplications, not taking into account the computation of the gradient $\nabla_{\mat{W}} R_{\mathrm{sum}}\left(\mat{W}^{(i)}\right)$. In order to compute the gradient, \cite{dab_mmwave} employs a finite differences approximation which is computed for each element as
\begin{align*}
    \left[\nabla_{\mat{W}} R_{\mathrm{sum}}(\mat{W})\right]_{m,k} \approx \frac{\left(R_{\mathrm{sum}}(\mat{W} + \mat{\Delta}^{(m,k)}_{\Re}) - R_{\mathrm{sum}}(\mat{W})\right) + j\left(R_{\mathrm{sum}}(\mat{W} + \mat{\Delta}^{(m,k)}_{\Im}) - R_{\mathrm{sum}}(\mat{W}) \right) }{\delta},
\end{align*}
where $\mat{\Delta}^{(m,k)}_{\Re} \in \mathbb{R}^{M\times K}$ is a matrix containing only zeros except for $\delta$ at row $m$ and column $k$, where $\delta$ is a small positive constant. Similarly $\mat{\Delta}^{(m,k)}_{\Re} \in \mathbb{R}^{M\times K}$ contains all zeros except for $j\delta$ at row $m$ and column $k$. The computation of the additions, subtractions and quotient in this equation requires $4$ real additions and $2$ real multiplications. Additionally, in this equation the sum rate needs to be evaluated 3 times according equations (\ref{eq:rate}), (\ref{eq:snidr_analytical}), (\ref{eq:gainmat}), (\ref{eq:cov_e}) and (\ref{eq:ln}).
Equation (\ref{eq:rate}) requires $K(K-1)$ complex additions, (\ref{eq:snidr_analytical}) requires $(K+1)(M^2-1)$ complex additions, $2$ real additions, $(K+1)(M^2-1) + 1$ complex multiplications and $1$ real multiplication. Equation (\ref{eq:gainmat}) requires $M^2(K-1)+5M$ complex additions and $M^2K + 9M$ complex multiplications, (\ref{eq:cov_e}) requires $5M^3 - 5M^2 +4M$ complex additions and $5M^3 + 15M$ multiplications and (\ref{eq:ln}) requires $10M$ additions and $10M$ multiplications. This leads to the following number of real floating point additions and multiplications respectively
\begin{align*}
     &P I \left[ 60M^4K^2 + 24M^3K^3 + 12M^3K^2 + 12M^2K^3 + 420 M^2K^2 + 15MK^2 + 4MK\right]  \\
     &P I \left[ 60M^4K^2 + 24M^3K^3 - 24M^3K^2 + 6M^2K^3 + 324 M^2K^2 - 3MK^3+ 3MK^2 + 6MK \right]
\end{align*}
This scales as $\mathcal{O}\left(P I M^4K^2 + P I M^3K^3  \right)$ real floating point additions, $\mathcal{O}\left(P I M^4K^2 + PI M^3K^3  \right) $ real floating point multiplications or $\mathcal{O}\left(P I M^4K^2 + P I M^3K^3  \right)$ \gls{flops}.

\subsection{Numerical Comparison}
An overview of the complexity of the \gls{gnn} and the different benchmarks is given in \cref{tab:complexity}. Additionally, a numerical example of the number of \gls{flops} is given for the following case: $P=50$, $I=1000$, $M=64$, $K=4$, $d_l=128$ and $L=8$. This illustrates that the \gls{gnn} precoder has a complexity that is six orders of magnitude smaller as compared to the \gls{dab} precoder. Moreover, when the ability of the \gls{gnn} to paralellize the computations across the edges is utilized, the number of \gls{flops} that need to be executed in serial is eight orders of magnitude smaller as compared to the \gls{dab} precoder. Next to this, we see that when the \gls{gnn} is parallelized, the number of serial \gls{flops} is approximately 200 times higher than the \gls{zf} precoder.

\begin{table}[t]
\centering
    \caption[Caption for LOF]{Complexity of the \gls{gnn}, \gls{zf} and \gls{dab} precoder. Example for the values $P=50$, $I=1000$, $M=64$, $K=4$, $d_l=128$ and $L=8$.}
    \label{tab:complexity}
    \centering

\makebox[\textwidth][c]{\resizebox{1\textwidth}{!}{
 \begin{tabular}{@{}cccccc@{}} 
 \toprule
     & \gls{gnn} & \gls{gnn} (Serial Operations) & \gls{dab} & \gls{zf}  \\ [0.5ex] 
     \midrule
 \arrayrulecolor{black!30}
   Additions & $\mathcal{O}\left(MKd_l^2L + M^2Kd_lL + MK^2d_lL \right)$ &  \ $\mathcal{O}\left(d_l^2L + Md_lL + Kd_lL \right)$ &  $\mathcal{O}\left(P I M^4K^2 + P I M^3K^3  \right)$ &  $\mathcal{O}(K^3 + K^2M)$ \\
\hline   

     Multiplcations & $\mathcal{O}\left(MKd_l^2L\right)$ &  \ $\mathcal{O}\left(d_l^2L \right)$ &  $\mathcal{O}\left(P I M^4K^2 + P I M^3K^3  \right)$ &  $\mathcal{O}(K^3 + K^2M)$ \\
\hline   

       FLOPs & $\mathcal{O}\left(MKd_l^2L + M^2Kd_lL + MK^2d_lL \right)$ &  \ $\mathcal{O}\left(d_l^2L + Md_lL + Kd_lL \right)$ &  $\mathcal{O}\left(P I M^4K^2 + P I M^3K^3  \right)$ &  $\mathcal{O}(K^3 + K^2M)$ \\
     
\hline   

    Example ($\sim$FLOPs) & \num{51}$\cdot 10^6$ & \num{20} $\cdot 10^4$ & \num{14}$\cdot 10^{12}$ & \num{1000}\\
    \arrayrulecolor{black}
  \bottomrule
\end{tabular}
}}
\end{table} 
To keep a concise and fair comparison between the GNN and the benchmarks, only the number of \gls{flops} is taken into account. However, specialized \gls{nn} accelerators can improve the inference times of \glspl{nn} and reduce the energy required to run them~\cite{nn_accel, hardware_survey}. Next to this a myriad of acceleration techniques can be applied to neural networks such as tensor decomposition, pruning, general matrix multiplication, Winograd transformation, etc.~\cite{nn_accel}. Additionally, \gls{nn}s are known to be able to work on a lower bit resolution, leading to the use of fixed point arithmetic rather than floating point arithmetic, which leads to even faster inference times~\cite{nn_accel}. Exploration of these avenues can lead to faster and more efficient execution of the \gls{gnn}.

\section{Simulation Results}\label{sec:results}
For the following simulations, the polynomial coefficients are obtained by a least squares regression of an $11^{\mathrm{th}}$ order model to the modified Rapp model as described in \cref{sec:pas}. This regression is performed at the desired \gls{ibo}, the polynomial coefficients corresponding to the \gls{ibo} value can be found in Appendix~\ref{ap:paparams}. For some specific cases a $3^{\mathrm{th}}$ order \gls{pa} model at an \gls{ibo} = $-3$ dB is considered, the PA parameters for this model are\footnote{Obtained using a least squares regression of the $3^{\mathrm{th}}$ order PA model to the modified Rapp model at an \gls{ibo} = \SI{-3}{\decibel}.} $\beta_1 = 1, \beta_3=-77.82-40.12j \cdot 10^{-3}$. For all simulations, the total transmit power is $P_T = M$. Hence, the average power at the input of each \gls{pa} is $p_{\mathrm{in}} = P_T/M = 1$. The linear \gls{pa} gain is set to one. Training and testing are done at an \gls{ibo} of \SI{-3}{\decibel}, unless specified otherwise, this saturates the \glspl{pa} significantly more than current cellular systems that require 9-12 \SI{}{\decibel} back-off. After training, the \glspl{nn} are evaluated based on the sum rate given in (\ref{eq:rate}) where the \gls{snidr} is computed numerically according to (\ref{eq:snidr_num}). The training set consists of \num{500000} generated channels. These are generally sampled from a complex normal distribution with zero mean and variance one $[\mat{H}]_{i,j} \sim \mathcal{CN}(0,1)$ if a Rayleigh fading channel is assumed. When considering radiation patterns, a deterministic \gls{los} channel is generated according to~(\ref{eq:los}), where the user angle (in degrees) is randomly sampled from a discrete uniform distribution $\theta_k \sim \mathcal{U}\{0, 180\}$. The hyperparameters of the neural networks are selected by using a validation set of size \num{2000}, while for the simulations performed in this section, an independent test set of size \num{10000} is used. For training, a batch size of 64 is used, with an initial learning rate of $5\times10^{-3}$, which is reduced if the validation loss reaches a plateau. The network is trained for 50 epochs with early stopping if the validation loss does not further decrease.

\subsection{Single-User Case} 
In this section the performance of the \gls{gnn} as described in \cref{sec:gnn_basic} is evaluated, for the single-user case $K=1$. This \gls{gnn} takes as input the channel matrix $\mat{H}$ and outputs the precoding matrix $\mat{\hat{W}}$. For the evaluations performed in this section, the \gls{gnn} consists of 8 layers (i.e., $L=8$), with each hidden layer having $d_l = 128$ features. Each \gls{gnn}-layer is described by (\ref{eq:updatebase}). Where the non-linear activation function in each layer is a \gls{lrelu}, except in the last \gls{gnn}-layer where a linear activation is used. After this linear activation, the power normalization layer as described in (\ref{eq:pwr_norm}) is performed.  During training, $P_T/\sigma^2_v$ is set to \SI{20}{\decibel}.

\begin{figure}[t]
    \centering
    \subfloat[$3^{\mathrm{rd}}$ order PA]{\label{fig:rsum_third_order_k1}\includegraphics{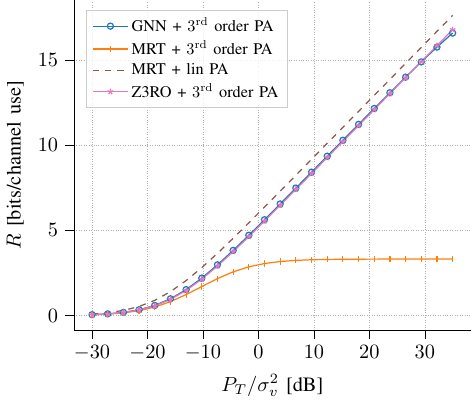}}
    \hfill
    \subfloat[$11^{\mathrm{th}}$ order PA]{\label{fig:rsumk1}\includegraphics{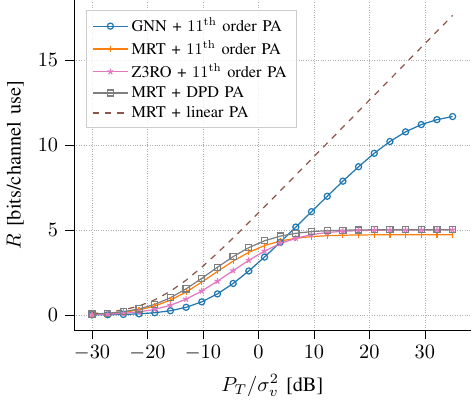}}
    \caption{Achievable rates averaged over the channel realizations taken from the test set evaluated on a $3^{\mathrm{rd}}$ order \gls{pa} (a) and an $11^{\mathrm{th}}$ order \gls{pa} (b). Comparing the \gls{gnn} precoder against MRT, MRT plus a perfect \gls{dpd} and the \gls{z3ro} precoder from~\cite{z3ro} for $M=64$ and $K=1$. Observations: (a) \gls{gnn} closely matches the optimal \gls{z3ro} precoder. (b) \gls{gnn} copes better with high-order effects in the \gls{pa}.}
    \label{fig:singleuserplots}
\end{figure}

\begin{figure}[tb]
    \centering
    \subfloat[GNN trained on $3^{\mathrm{rd}}$ order PA]{\label{fig:third_order_amp_phase}\includegraphics[]{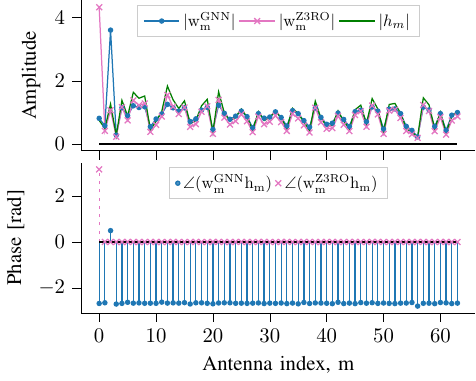}}\hfill
    \subfloat[GNN trained on $11^{\mathrm{th}}$order PA]{\label{fig:11_order_amp_phase}\includegraphics[]{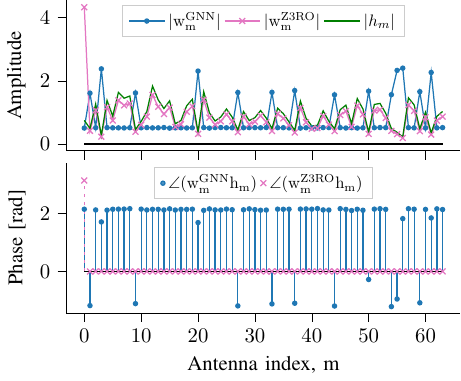}}
    \caption{Amplitude and phase plot per antenna, \gls{gnn} compared to the \gls{z3ro} precoder from~\cite{z3ro}. When trained on the third-order model (a), the \gls{gnn} replicates the behavior of the optimal Z3RO precoder i.e., saturate one antenna with an opposite phase shift to cancel the third order distortion. When trained on an eleventh order PA (b), the \gls{gnn} saturates more antennas in order to spatially suppress the high-order distortion.}
    \label{fig:amp_phase_plots}
\end{figure}

For the single-user case when one assumes the simplified third-order polynomial \gls{pa} model (i.e., $N=1$ in (\ref{eq:poly})) a closed-form solution exists, namely the \gls{z3ro} precoder. Therefore, first it is investigated whether the \gls{gnn} precoder can achieve the same performance as the optimal \gls{z3ro} precoder under these assumptions. In \cref{fig:rsum_third_order_k1} the achievable rates of  the \gls{gnn}, \gls{z3ro} and \gls{mrt} precoder are depicted as a function of $P_T/\sigma^2_v$. The \gls{mrt} precoder is designed under the assumption of a linear \gls{pa}. As a consequence, when a non-linear \gls{pa} is present, the user performance degrades as compared to the linear case, as can be seen in \cref{fig:rsum_third_order_k1}. Additionally, it can be seen in \cref{fig:rsum_third_order_k1} that the rate of the \gls{z3ro} precoder is not limited by distortion but grows logarithmically with $P_T/\sigma^2_v$. Indeed, the \gls{z3ro} precoder is able to  mitigate the third-order distortion in the user direction by saturating one (or a few) antennas with an opposite phase shift, which comes at the cost of a small reduction in array gain. When comparing the \gls{gnn} precoder, trained on a third-order \gls{pa} model, against the \gls{z3ro} precoder we see that the \gls{gnn} achieves similar performance as the \gls{z3ro} precoder. Indeed, when comparing the amplitude and phase of both precoders in \cref{fig:third_order_amp_phase}, it is clear that the \gls{gnn} also saturates one or a few of the antennas with an opposite phase shift. In conclusion, for $K=1$, the \gls{gnn} has learned a similar precoding structure as the optimal \gls{z3ro} precoder.

Second, the case of a high-order \gls{pa} model is studied, namely an eleventh order \gls{pa} ($N=5$). In this case the \gls{z3ro} precoder is no longer optimal as can be seen in \cref{fig:rsumk1}. Under this \gls{pa} model the rate of the \gls{z3ro} precoder is limited by distortion as $P_T/\sigma^2_v$ grows, this is also the case for the \gls{mrt} precoder even when it is combined with an ideal \gls{dpd}. When comparing against the \gls{gnn} precoder, trained on an eleventh order \gls{pa}, \cref{fig:rsumk1} illustrates how the \gls{gnn} drastically outperforms \gls{mrt}, \gls{z3ro} precoding and even \gls{mrt} coupled with a perfect \gls{dpd}. The \gls{gnn} provides a rate which nearly grows logarithmically with  $P_T/\sigma^2_v$ and is only limited by distortion for very large  $P_T/\sigma^2_v$. Note that the \gls{gnn} precoder only outperforms the benchmarks for large $P_T/\sigma_{\nu}^2$ i.e., when the system is mostly corrupted by distortion and not by noise.

\begin{figure}[t!]
    \centering
    \subfloat[$P_{\mathrm{lin}}(\Tilde{\theta})$, $
P_{\mathrm{dist}}(\Tilde{\theta})$ for MRT]{\label{fig:rad_k1_mrt}\includegraphics[]{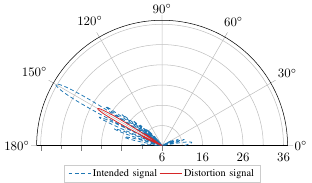}}\hfill
    \subfloat[$P_{\mathrm{lin}}(\Tilde{\theta})$, $
P_{\mathrm{dist}}(\Tilde{\theta})$ for GNN]{\label{fig:rad_k1_gnn}\includegraphics[]{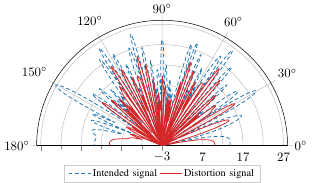}}\hfill
     \subfloat[$P_{\mathrm{SDR}}(\Tilde{\theta})$ for MRT and GNN]{\label{fig:rad_k1_sdr}\includegraphics[]{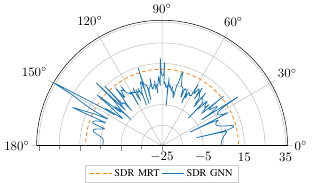}}
	\caption{Radiation pattern of the intended signal $P_{\mathrm{lin}}(\Tilde{\theta})$ and distortion signal $
P_{\mathrm{dist}}(\Tilde{\theta})$ [dB] for an $11^{\mathrm{th}}$ order PA, $K=1$ of the \gls{gnn} precoder (a) and \gls{zf} precoder (b) for a pure \gls{los} channel and a half-wavelength \gls{ula}. User angle is $\theta=150^{\circ}$. In (c) the \acrfull{sdr} radiation pattern $P_{\mathrm{SDR}}(\Tilde{\theta})$ is depicted [dB]. }
	\label{fig:rad_k1} 
\end{figure}

When looking at the amplitude and phase of the \gls{gnn} precoder trained on the $11^{\mathrm{th}}$ order PA in~\cref{fig:11_order_amp_phase} it is clear that it no longer behaves as the \gls{z3ro} precoder (i.e., perform MRT precoding with one or a few saturated antennas with opposite phase shift to cancel distortion). Rather, it keeps the amplitude of nearly all transmit antennas constant except for a number of antennas which are saturated with a different phase shift. This avoids sending all distortion in the user direction as can be seen in the radiation patterns in \cref{fig:rad_k1}. In \cref{fig:rad_k1_mrt}, the radiation pattern of the intended and distortion singal are depicted for the \gls{mrt} precoder, here it is clear that the distortion is beamformed in the same direction as the intended signel, i.e., the user direction. \Cref{fig:rad_k1_gnn}, depicts the radiation patterns of the intended and distorted signals for the proposed \gls{gnn} precoder. This figure illustrates that the distortion is more uniformly distributed for the \gls{gnn} precoder and even close to zero in the user direction. The cost of this dispersed distortion is a reduction in array gain in the user direction. However, as \cref{fig:rad_k1_sdr} illustrates, the \gls{sdr} radiation pattern for the \gls{gnn} precoder has a clear peak in the user direction, while it is uniform for the \gls{mrt} precoder. 

\subsection{Multi-User Case}
In this section, the same \gls{gnn} model as in the previous section is used, however, this time the multi-user case is considered.
When multiple users are present, both the inter-user interference and distortion to multiple users have to be mitigated. In this more complex scenario, there is no closed-form solution available for the optimal precoder. As such, the \gls{gnn} is trained to learn how to perform this task. Additionally, the \gls{dab} precoder from~\cite{dab_mmwave} is considered, as it provides a way to solve the non-convex optimization problem, without any guarantees on optimality. Note that in~\cite{dab_mmwave} the optimization procedure is repeated for $P=50$ different initializations and executed for $I=100$ iterations per initialization. In~\cite{dab_mmwave} this procedure is reported to converge after $I=100$ iterations. However, given the added complexity of our scenario (i.e., 64 vs. 16 Tx antennas and an $11^{\mathrm{th}}$-order vs. a $5^{\mathrm{th}}$-order \gls{pa}) our simulations did not converge after $I=100$ iterations. As such, the simulation results for the \gls{dab} precoder in this study are obtained by running the optimization procedure for $I=1000$ iterations, repeated for $P=50$ different initializations. In order to compensate for this added simulation time the \gls{dab} precoder is only evaluated on 10 channel realizations taken from the test set, while the \gls{gnn} is evaluated on the full test set\footnote{The results when evaluating the GNN on the same 10 channel realizations as the ones used to evaluate the DAB precoder match the results obtained when evaluating the GNN on the full test set.}. %

\begin{figure}[t]
    \centering
    \subfloat[]{\label{fig:rsumk24}\includegraphics[]{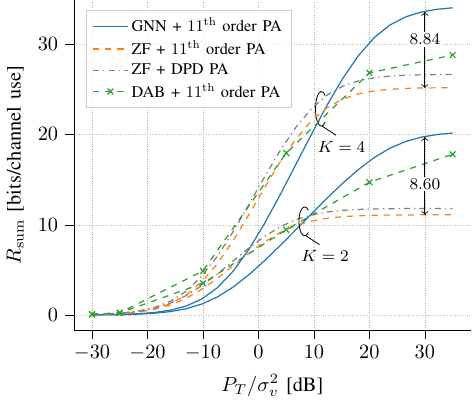}}
    \hfill
    \subfloat[]{\label{fig:rsum_vs_ibo}\includegraphics[]{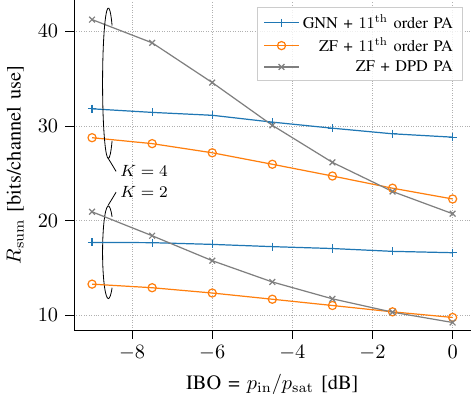}}
    \caption{Achievable rates averaged over the channel realizations taken from the test set evaluated on an $11^{\mathrm{th}}$ order \gls{pa}. Comparing the \gls{gnn} precoder against \gls{zf} and the \gls{dab} precoder from~\cite{dab_mmwave} for $M=64$ and $K\in \{2,4\}$ at $\mathrm{IBO}=-3$ dB in (a). Note that the DAB precoder was only evaluated on 10 channel realizations to reduce simulation time. In (b) the achievable sum rates are plotted for $M=64$, $K\in \{2,4\}$, $P_T/\sigma_{\nu}^2 = 20$ dB with varied \gls{ibo}. Comparing the \gls{gnn}, \gls{zf} and \gls{zf} plus \gls{dpd}. The \gls{gnn} is retrained at each IBO point.
 }
    \label{fig:rsumk24_plus_ibo}
\end{figure}

In \cref{fig:rsumk24}, a comparison between the \gls{gnn}, \gls{dab}, \gls{zf} and \gls{zf} precoder coupled with a perfect \gls{dpd} is made. In this figure, the sum rate is depicted in function of $P_T/\sigma^2_{\nu}$ for $K \in \{2, 4\}$, when using the eleventh order \gls{pa} model. Comparing the \gls{gnn} precoder against the \gls{zf} precoder, for $K=2$ the sum rate is increased by 8.60 bits/channel use at $P_T/\sigma^2_{\nu} = 30$ dB, when using the proposed method, and 8.84 bits/channel use for $K=4$. The \gls{gnn} precoder shows superior performance over the \gls{dab} precoder. Next to this, it is shown that the proposed solution outperforms a perfect \gls{dpd} combined with \gls{zf}, when the system is highly distortion limited (i.e., for high $P_T/\sigma^2_{\nu}$ levels). This shows the ability of the \gls{gnn} to cancel high-order non-linear distortion in the multi-user scenario, which results in significant increases in capacity. Additionally, this illustrates that the higher the number of users becomes, the less gain is to be obtained, in terms of added rate per user, by using the \gls{gnn} precoder. This is expected as non-linear distortion is more spatially spread out when more users are present~\cite{distortion_beamformed2}. In other words, less distortion is beamformed in the user directions, which leads to less potential gains for mitigating this distortion. Moreover, when more users are present, canceling all distortion to all users becomes more complex. Nevertheless, the \gls{gnn}-based precoder achieves a significant increase in achievable sum rate as compared to the benchmark approaches. 

\Cref{fig:rsum_vs_ibo} depicts the sum rate when $p_{\mathrm{in}}$ is fixed but $p_{\mathrm{sat}}$ is varied, resulting in a varied \gls{ibo}. This shows that, for $K\in\{2,4\}$, the \gls{gnn} always outperforms the classical \gls{zf} precoder, when evaluated using the eleventh order \gls{pa} model. Moreover, when $K=2$, the \gls{gnn} is able to achieve a nearly constant sum rate over a wide range of \gls{ibo}. This illustrates the ability of the \gls{gnn} to suppress nearly all high-order distortion in the user directions. When comparing the \gls{gnn} to \gls{zf} plus a perfect \gls{dpd} in \cref{fig:rsum_vs_ibo}, it is evident that the \gls{gnn} is most beneficial when a lot of distortion is present, i.e., at low back-off. However, we stress the fact that the \gls{gnn} does not have to be used as a replacement, but could be used in combination with \gls{dpd}. When (perfect) \gls{dpd} is available, the \gls{pa} characteristic after applying \gls{dpd} can be modeled as a polynomial on which the \gls{gnn} can be retrained. As such, the combination of both approaches could produce even better sum rates. However, this would add a significant complexity burden.

\begin{figure}[t!]
    \centering
    \subfloat[$P_{\mathrm{lin}}(\Tilde{\theta})$, $P_{\mathrm{dist}}(\Tilde{\theta})$ for ZF]{\label{fig:rad_k2_zf}\includegraphics[]{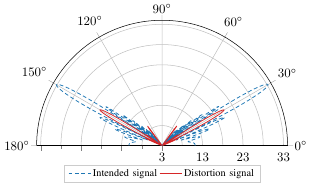}}\hfill
    \subfloat[$P_{\mathrm{lin}}(\Tilde{\theta})$, $P_{\mathrm{dist}}(\Tilde{\theta})$ for GNN]{\label{fig:rad_k2_gnn}\includegraphics[]{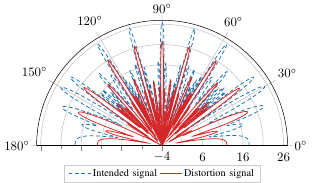}}\hfill
     \subfloat[$P_{\mathrm{SDR}}(\Tilde{\theta})$ for ZF and GNN]{\label{fig:rad_k2_sdr}\includegraphics[]{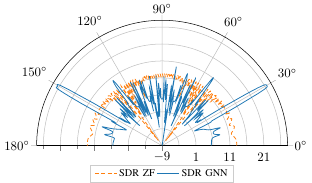}}
	\caption{Radiation pattern of the intended signal $P_{\mathrm{lin}}(\Tilde{\theta})$ and distortion signal $
P_{\mathrm{dist}}(\Tilde{\theta})$ [dB] for an $11^{\mathrm{th}}$ order PA, $K=2$ of the \gls{gnn} precoder (a) and \gls{zf} precoder (b) for a pure \gls{los} channel and a half-wavelength \gls{ula}. User angles are $\theta\in \{30^{\circ}, 150^{\circ}\}$. In (c) the \acrfull{sdr} radiation pattern $P_{\mathrm{SDR}}(\Tilde{\theta})$ is depicted [dB].}
	\label{fig:rad_k2} 
\end{figure}

\begin{figure}[tt!]
    \centering
    \subfloat[$P_{\mathrm{lin}}(\Tilde{\theta})$, $P_{\mathrm{dist}}(\Tilde{\theta})$ for ZF]{\label{fig:rad_k4_zf}\includegraphics[]{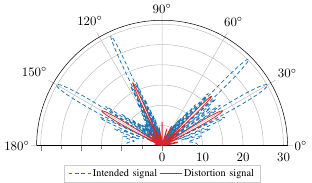}}\hfill
    \subfloat[$P_{\mathrm{lin}}(\Tilde{\theta})$, $P_{\mathrm{dist}}(\Tilde{\theta})$ for GNN]{\label{fig:rad_k4_gnn}\includegraphics[]{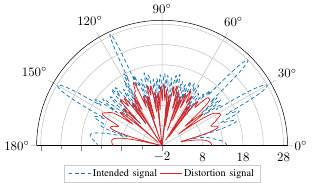}}\hfill
     \subfloat[$P_{\mathrm{SDR}}(\Tilde{\theta})$ for ZF and GNN]{\label{fig:rad_k4_sdr}\includegraphics[]{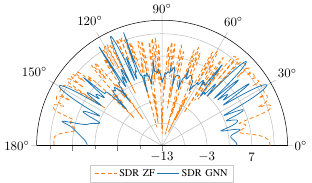}}
	\caption{Radiation pattern of the intended and distortion signal $P_{\mathrm{lin}}(\Tilde{\theta})$, $
P_{\mathrm{dist}}(\Tilde{\theta})$ [dB] for an $11^{\mathrm{th}}$ order PA, $K=4$ of the \gls{gnn} (a) and \gls{zf} precoder (b) for a pure \gls{los} channel and a half-wavelength \gls{ula}. User angles are $\theta\in \{30^{\circ}, 45^{\circ}, 115^{\circ},  150^{\circ}\}$. In (c) the \acrfull{sdr} radiation pattern $P_{\mathrm{SDR}}(\Tilde{\theta})$ is depicted [dB].}
	\label{fig:rad_k4} 
\end{figure}

Finally, in \cref{fig:rad_k2,fig:rad_k4} the radiation patterns for $K=2, 4$ are depicted. In \cref{fig:rad_k2_zf,fig:rad_k2_gnn}, the radiation patterns of the intended and distorted signals are plotted for the \gls{zf} and \gls{gnn} precoder respectively, for $K=2$. \Cref{fig:rad_k2_zf} clearly illustrates how for \gls{zf} precoding the distortion is beamformed in both user directions. \Cref{fig:rad_k2_gnn} reveals that for \gls{gnn} precoding the distortion is mainly beamformed in the non-user directions, as the distortion in the user directions is very small, this however comes at the cost of a reduction in array gain for the linearly amplified signal. Nonetheless, in \cref{fig:rad_k2_sdr} the \gls{sdr} radiation pattern is presented, which shows two clear peaks in the user directions, for the \gls{gnn} precoder, while the \gls{sdr} for the \gls{zf} precoder is nearly uniformly distributed. In \cref{fig:rad_k4_zf,fig:rad_k4_gnn}, the radiation patterns of the intended and distorted signals are plotted for the \gls{zf} and \gls{gnn} precoder respectively, for $K=4$ users. For the \gls{zf} precoder, the distortion is again beamformed in the user directions while it is more spatially spread out for the \gls{gnn} precoder. When looking at the \gls{sdr} radiation pattern in \cref{fig:rad_k4_sdr}, we see a slight improvement of the \gls{sdr} in the user directions for the \gls{gnn} precoder as compared to the \gls{zf} precoder.

\subsection{Improvements in Power Consumption}
\subsubsection{PA Power Consumption}
\begin{figure}[t]
    \centering
    \subfloat[]{\label{fig:pcons}\includegraphics[]{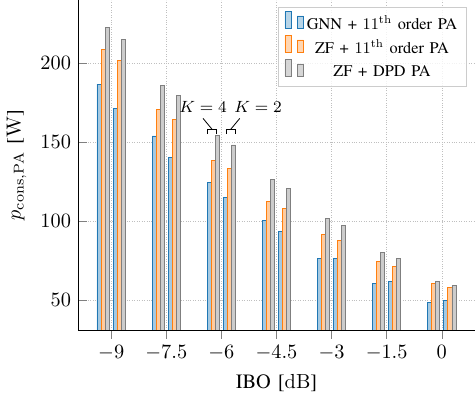}}
    \hfill
    \subfloat[]{\label{fig:rsum_vs_pcons}\includegraphics[]{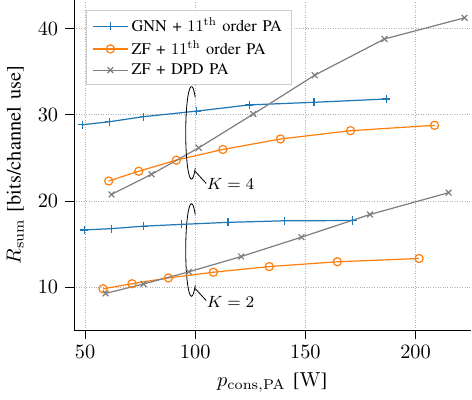}}
    \caption{(a) Consumed \gls{pa} power in function of \gls{ibo} (b) achievable sum rate in function of consumed \gls{pa} power.
    Both (a) and (b) compare the \gls{gnn}, \gls{zf} and \gls{zf} + DPD precoders for $M=64$, $K\in\{2, 4\}$, $\frac{P_T}{\sigma^2_{\nu}} = 20$ dB.}
    \label{fig:pcons_gains}
\end{figure}

By spatially suppressing the non-linear distortion, a lower \gls{ibo} can be used to achieve a desired sum rate. By allowing the \glspl{pa} to work closer to saturation, their energy efficiency is improved. In this section the improvements in \gls{pa} power consumption are quantified. For a class B amplifier the following model of the energy efficiency can be used~\cite{amp_eff}. The efficiency of PA $m$ is defined as
\begin{align}
    \eta_m=\eta_{\max } \sqrt{\frac{p_m}{p_{\mathrm{sat} }}}. \label{eq:etam}
\end{align}
Here $p_{\mathrm{sat}}$ is the saturation power of the \gls{pa}, while $\eta_{\mathrm{max}}$ is the maximum achievable efficiency, for a class B amplifier $\eta_{\mathrm{max}}=0.785$~\cite{amp_eff}. Next to this, $p_m$ is defined as the power at the output of amplifier $m$
\begin{align}
    p_m = \expt{|y_m|^2} = \expt{|\phi(x_m)|^2} = \expt{\left|\phi\left(\sum_{k=0}^{K-1}w_{m,k}s_k\right)\right|^2}. \label{eq:pm}
\end{align}
The total consumed \gls{pa} power is then given by
\begin{align}
    p_{\mathrm{cons}, \mathrm{PA}} = \sum_{m=0}^{M-1} \frac{p_m}{\eta_m} =  \frac{\sqrt{p_{\mathrm{sat}}}}{\eta_{\mathrm{max}}}\sum_{m=0}^{M-1} \sqrt{ \expt{|y_m|^2}}.
\end{align}
From this expression, it can be seen that the consumed power is not proportional to the average input back-off but rather to the sum of the square root of the output power of the \glspl{pa}. In \cref{fig:pcons} the consumed power of the PAs in function of the input back-off is depicted for $K=2$ and $K=4$. This figure illustrates that the \gls{gnn} has a lower PA consumed power as compared to the \gls{zf} and \gls{zf} + \gls{dpd} precoding, for a same average input back-off (and thus a same transmit power). Intuitively, this lower consumption is due to the fact that the \gls{gnn} saturates a number of antennas. These saturated antennas have a very high efficiency, while they also receive more power as compared to the less efficient antennas.

Moreover, for a fair comparison one needs to select a quality of service constraint, e.g., a sum rate constraint, and compare the consumed power for the different precoders meeting this rate constraint. As an example for $M=64$, $K=4$ let us consider a desired sum rate of 30 bits/channel use. From \cref{fig:rsum_vs_pcons} the consumed power to achieve this rate can be seen, for GNN precoding this is $p_{\mathrm{cons}, \mathrm{PA}} =$ \SI{85.57}{\watt}, for ZF precoding $p_{\mathrm{cons}, \mathrm{PA}} =$ \SI{284.29}{\watt} and for ZF precoding + DPD this is $p_{\mathrm{cons}, \mathrm{PA}} =$ \SI{125.98}{\watt}.\footnote{Obtained using linear interpolation between the points of \cref{fig:rsum_vs_pcons}.} This reveals that the \gls{gnn} precoder has a PA power consumption that is 3.32 times smaller as compared to \gls{zf} precoding and 1.47 times smaller as compared to \gls{zf} precoding plus \gls{dpd}.

\subsubsection{Digital Signal Processing Power Consumption}
In this section the consumed power of the hardware accelerator needed to run the \gls{gnn} is quantified. Considering the scenario were  $M=64$, $K=4$, based on \cref{sec:complexity} it is clear that 164 M\gls{flops} are required to compute one forward pass of the \gls{gnn}. In order to establish the desired speed (\gls{flops}/s) of the hardware accelerator, the coherence time of the channel is computed to determine the time available to compute the precoding coefficients. The coherence time is proportional to the inverse of the maximum Doppler frequency $f_m = \frac{v}{c} f_c$ and is often defined as $T_c = \frac{1}{2 f_m}$~\cite{fundamentals_mimo}. Assuming a carrier frequency of $f_c= 5$ GHz and a receiver velocity of $v=10$ m/s, this gives a coherence time of \SI{3}{\milli\second}. As a rule of thumb it is assumed that 10\% of the coherence time can be used to compute the precoding coefficients. This establishes a required hardware accelerator speed of 549 \gls{gops}. Based on~\cite{hardware_survey} the state-of-the-art Manticore accelerator~\cite{manticore} is selected, which can achieve the desired speed and precision.\footnote{Note that the \gls{gnn} considered in this work requires single precision (32 bit) floating point operations.} The full Manticore design consists out of 4096 cores, however in~\cite{manticore} a prototype with 24 cores was developed and measured. This prototype is able to achieve 50 \gls{gops} with a power consumption of \SI{200}{\milli\watt}. Given the parallel architecture of the \gls{gnn} it can be deployed across multiple cores. Hence in order to reach the desired speed of 549 \gls{gops} 264 cores can be utilized rather than 24, this leads to a total \gls{dsp} power consumption of \SI{2.2}{\watt}.

\subsubsection{Total Power Consumption}
An overview of the total consumed power (\gls{pa} and processing power) for the \gls{gnn} can be seen in \cref{tab:totalgain}. From this table it can be seen that the total consumed power for GNN precoding is \SI{87.77}{\watt}, for ZF precoding \SI{284.29}{\watt} and for ZF + DPD \SI{125.98}{\watt}. Note that as a worst-case scenario the processing power of \gls{zf} and \gls{dpd} is assumed to be zero, while in reality this processing power will be substantial when \gls{dpd} is implemented. Even in this scenario, the GNN precoder consumes 3.24 times less power as compared to ZF and 1.44 times less power as compared to ZF+DPD.

\begin{table}[b]
\centering
    \caption[Caption for LOF]{Consumed power of the \gls{gnn}, \gls{zf} and \gls{zf} + DPD precoders for $M=64$, $K=4$, for a fixed data rate of 30 bits/channel use. As a worst-case scenario the processing power for ZF and DPD are assumed to be zero.}
    \centering
 \begin{tabular}{@{}ccccc@{}} 
 \toprule
      Precoder & Precision  & Accelerator power [W]  & $p_{PA, cons}$ [W] & total power [W]\\ [0.5ex] 
\midrule
 \arrayrulecolor{black!30}
    GNN + Manticore~\cite{manticore}& \gls{sp} & 2.2 &  85.57 & 87.77  \\
\hline   

       ZF + DPD & \textit{NA} & \textit{NA} &  125.98 &  125.98  \\
\hline   
       ZF  & \textit{NA} &\textit{NA} &  284.29 &   284.29  \\
    \arrayrulecolor{black}
\bottomrule
\end{tabular}\label{tab:totalgain}
\end{table}

\subsection{Validation on Rapp PA Model}
\begin{figure}[t]
    \centering
    \includegraphics[]{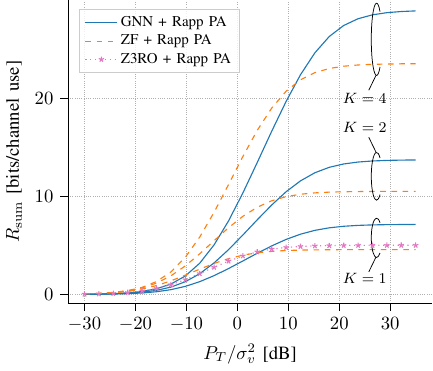}
    \caption{Achievable sum rates averaged over 1000 Rayleigh channel realizations taken from the test
set evaluated using the Rapp \gls{pa} model (\ref{eq:amamrapp}, \ref{eq:ampmrapp}). $M=64$, $K\in \{1,2,4\}$, with a varied $P_T/\sigma_{\nu}^2$.}
    \label{fig:rsumrapp}
\end{figure}

In this section, the \gls{gnn} precoder which is trained on the eleventh order \gls{pa} model is evaluated on the modified Rapp model. The modified Rapp model is originally used to obtain the coefficients of the polynomial model by performing a least squares regression, as discussed in \cref{sec:pas}. By evaluating the \gls{gnn} on the modified Rapp model, the effects of using a polynomial model, rather than the actual \gls{pa} characteristic can be studied. 
In \cref{fig:rsumrapp} the rate is depicted in function of $P_T/\sigma^2_{\nu}$ when the modified Rapp model is used for the amplification stage. For the single-user case $K=1$, this figure illustrates how the \gls{gnn} precoder trained on a polynomial \gls{pa} still outperforms the \gls{mrt} and \gls{z3ro} precoders when the underlying modified Rapp model is assumed. For the multi-user case $K\in \{2, 4\}$ in \cref{fig:rsumrapp}, it is again clear that the \gls{gnn} precoder outperforms the \gls{zf} precoder when the modified Rapp model is deployed as the amplification model. This highlights the fact that the polynomial model can successfully be used to train the \gls{gnn} even if the underliying PA characteristic is slightly different. 

 \subsection{Leveraging SNR Regime Information}
In this section the performance of the \gls{gnn} as described in \cref{sec:snrgnn} is evaluated. This \gls{gnn} takes as input edge features the channel coefficients and the normalized SNR, and is denoted as GNN\textsubscript{SNR}. GNN\textsubscript{SNR} is compared against GNN\textsubscript{base} which only takes the channel matrix as input. The \gls{gnn} architecture is kept the same as in \cref{sec:gnn_basic} except for the case where $K=4$ where $d_l=256$ features per layer are used rather than 128. 

\begin{figure}[t]
    \centering
    \subfloat[$K=1$]{\label{fig:snrk1}\includegraphics[]{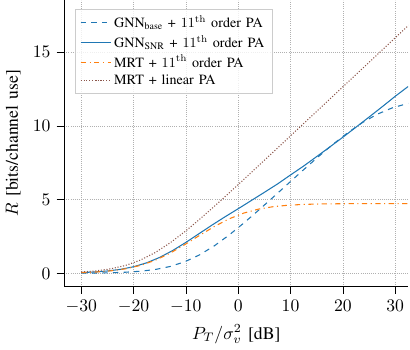}}
    \hfill
    \subfloat[$K\in\{2, 4\}$]{\label{fig:snrk24}\includegraphics[]{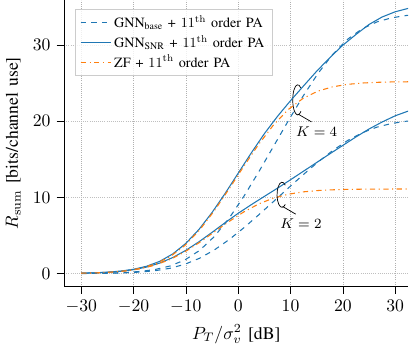}}
    \caption{Achievable rates for $M=64$, $K\in\{1, 2, 4\}$ and an $11^{\mathrm{th}}$order PA. Comparing the base GNN from \cref{sec:gnn_basic} against the GNN that leverages SNR information from \cref{sec:snrgnn}.}
    \label{fig:snrgnn}
\end{figure}

In \cref{fig:snrk1} and~\ref{fig:snrk24} the sum rate is depicted in function of $P_T/\sigma^2_{\nu}$ for $K \in \{1, 2, 4\}$, for the eleventh order \gls{pa} model.  This figure illustrates the ability of GNN\textsubscript{SNR} to leverage the extra SNR input to deliver performance over the full range of $P_T / \sigma^2_{\nu}$. This alleviates the need for retraining the \gls{gnn} at each value of $P_T / \sigma^2_{\nu}$. In this way, GNN\textsubscript{SNR} not only performs well in the distortion limited regime (i.e., for high $P_T / \sigma^2_{\nu}$) but is also capable of delivering the same performance as \gls{zf} in the noise limited regime (i.e., for low $P_T / \sigma^2_{\nu}$), which GNN\textsubscript{base} was not able to do.

\FloatBarrier
\section{Conclusion}
\label{sec:conclusion}
In this study, a \gls{mmimo} system is considered, where a \gls{gnn} is trained to learn a mapping between the channel and precoding matrix which maximizes the sum rate in the presence of high-order non-linear \gls{pa} distortion. The complexity of the proposed solution is analysed and shown to be six orders of magnitude lower as compared to the benchmark \gls{dab} precoder. Simulation results indicate that the proposed solution offers an increased capacity as compared to \gls{zf} in the distortion-limited regime, i.e., an energy-efficient regime. For instance, at an \gls{ibo} of \SI{-3}{\decibel} the proposed precoder produces an increase in sum rate of 8.60 and 8.84 bits/channel use for the two and four user cases respectively. Additionally, it is shown that the proposed solution achieves a higher sum rate as compared to \gls{zf} precoding plus \gls{dpd}, as \gls{dpd} is limited by the distortion caused by saturation based clipping. Radiation patterns demonstrate that the higher rate is achieved by spatially suppressing the distortion in the user-directions. By utilizing the proposed precoder, \glspl{pa} can operate closer to saturation, which increases their energy efficiency. For instance, it is shown that, in the four user-case, for a fixed sum rate the PA consumed power of the proposed precoder is 3.32 times and 1.47 times lower as compared to \gls{zf} precoding and \gls{zf} plus \gls{dpd} respectively. In addition, it is shown that the PA power consumption reduction is achieved at the price of a small increase in \gls{dsp} consumption so that the total power consumption is still greatly reduced. Furthermore, the \gls{gnn} is extended to incorporate additional knowledge, i.e., the \gls{snr} regime, which alleviates the need to retrain at each \gls{snr} point. These results are especially promising in the multi-user case where no closed-form/low-complexity solution to the distortion-aware precoding problem is available. 

Future perspectives include the incorporation of additional knowledge into the precoder design, such as the \gls{pa} parameters, which would allow the precoder to anticipate long-term changes in the \gls{pa} characteristics. Additionally, more complex \gls{pa} models could be investigated that capture memory effects. Finally, the impact of channel estimation and \gls{pa} parameter estimation errors on the performance of the proposed method should be evaluated in future studies.

\bibliographystyle{IEEEtran}
\bibliography{References}

% Generated by IEEEtran.bst, version: 1.14 (2015/08/26)
\begin{thebibliography}{10}
\providecommand{\url}[1]{#1}
\csname url@samestyle\endcsname
\providecommand{\newblock}{\relax}
\providecommand{\bibinfo}[2]{#2}
\providecommand{\BIBentrySTDinterwordspacing}{\spaceskip=0pt\relax}
\providecommand{\BIBentryALTinterwordstretchfactor}{4}
\providecommand{\BIBentryALTinterwordspacing}{\spaceskip=\fontdimen2\font plus
\BIBentryALTinterwordstretchfactor\fontdimen3\font minus \fontdimen4\font\relax}
\providecommand{\BIBforeignlanguage}[2]{{%
\expandafter\ifx\csname l@#1\endcsname\relax
\typeout{** WARNING: IEEEtran.bst: No hyphenation pattern has been}%
\typeout{** loaded for the language `#1'. Using the pattern for}%
\typeout{** the default language instead.}%
\else
\language=\csname l@#1\endcsname
\fi
#2}}
\providecommand{\BIBdecl}{\relax}
\BIBdecl

\bibitem{trends2040}
L.~Belkhir and A.~Elmeligi, ``\BIBforeignlanguage{en}{{Assessing ICT global emissions footprint: Trends to 2040 \& recommendations}},'' \emph{\BIBforeignlanguage{en}{Journal of Cleaner Production}}, vol. 177, pp. 448--463, Mar. 2018.

\bibitem{real_climate_impact_ict}
C.~Freitag, M.~Berners-Lee, K.~Widdicks, B.~Knowles, G.~S. Blair, and A.~Friday, ``{The real climate and transformative impact of ICT: A critique of estimates, trends, and regulations},'' \emph{Patterns}, vol.~2, no.~9, p. 100340, 2021.

\bibitem{greendeal}
{European Commission}, ``{The European Green Deal},'' \emph{COM (2019)}, November 2019.

\bibitem{sdgs}
{United Nations}, ``{The 2030 Agenda and the Sustainable Development Goals: An opportunity for Latin America and the Caribbean}.''

\bibitem{paris_agreement}
\BIBentryALTinterwordspacing
{United Nations Environment Programme}, ``{Paris Agreement},'' 12/12/2015. [Online]. Available: \url{https://wedocs.unep.org/20.500.11822/20830}
\BIBentrySTDinterwordspacing

\bibitem{ericsonmobility}
``{Ericsson Mobility report November 2022},'' \url{https://www.ericsson.com/4ae28d/assets/local/reports-papers/mobility-report/documents/2022/ericsson-mobility-report-november-2022.pdf}.

\bibitem{towards_green_5g}
L.~M.~P. Larsen, H.~L. Christiansen, S.~Ruepp, and M.~S. Berger, ``{Toward Greener 5G and Beyond Radio Access Networks--A Survey},'' \emph{IEEE Open Journal of the Communications Society}, vol.~4, pp. 768--797, 2023.

\bibitem{energy_performance_ericsson}
C.~Andersson, J.~Bengtsson, G.~Bystr\"{o}m, P.~Frenger, Y.~Jading, and M.~Nordenstr\"{o}m, ``{Improving energy performance in 5G networks and beyond},'' \emph{Ericsson Technology Review}, vol. 2022, no.~8, pp. 2--11, 2022.

\bibitem{how_much_energy}
G.~Auer, V.~Giannini, C.~Desset, I.~Godor, P.~Skillermark, M.~Olsson, M.~A. Imran, D.~Sabella, M.~J. Gonzalez, O.~Blume, and A.~Fehske, ``\BIBforeignlanguage{eng}{{How much energy is needed to run a wireless network?}}'' \emph{\BIBforeignlanguage{eng}{IEEE wireless communications}}, vol.~18, no.~5, pp. 40--49, 2011.

\bibitem{huawei_pa}
\BIBentryALTinterwordspacing
C.~Gabriel, A.~Hansang, and A.~Chern, ``{Green 5G: Building A Sustainable World},'' \emph{{Huawei Technologies Co.}}, 2020. [Online]. Available: \url{https://www.huawei.com/en/public-policy/green-5g-building-a-sustainable-world}
\BIBentrySTDinterwordspacing

\bibitem{rf_imperfections_pas}
T.~Schenk, \emph{\BIBforeignlanguage{eng}{{RF Imperfections in High-rate Wireless Systems: Impact and Digital Compensation}}}.\hskip 1em plus 0.5em minus 0.4em\relax Dordrecht: Springer Netherlands, 2008.

\bibitem{mmwave_nonlinearities}
N.~N. Moghadam, G.~Fodor, M.~Bengtsson, and D.~J. Love, ``{On the Energy Efficiency of MIMO Hybrid Beamforming for Millimeter-Wave Systems With Nonlinear Power Amplifiers},'' \emph{IEEE Transactions on Wireless Communications}, vol.~17, no.~11, pp. 7208--7221, 2018.

\bibitem{Eriksson2019NonlinearEO}
\BIBentryALTinterwordspacing
T.~Eriksson, W.~Cao, and C.~Fager, \emph{{Nonlinear Effects of Wireless Transceivers}}.\hskip 1em plus 0.5em minus 0.4em\relax John Wiley \& Sons, Ltd, 2019, pp. 1--30. [Online]. Available: \url{https://onlinelibrary.wiley.com/doi/abs/10.1002/9781119471509.w5GRef009}
\BIBentrySTDinterwordspacing

\bibitem{powerconsumption_dpd}
J.~Zanen, E.~Klumperink, and B.~Nauta, ``{Power Efficiency Model for MIMO Transmitters Including Memory Polynomial Digital Predistortion},'' \emph{IEEE Transactions on Circuits and Systems II: Express Briefs}, vol.~68, no.~4, pp. 1183--1187, 2021.

\bibitem{papr_reduction}
P.~P. Ann and R.~Jose, ``{Comparison of PAPR reduction techniques in OFDM systems},'' in \emph{2016 International Conference on Communication and Electronics Systems (ICCES)}, 2016, pp. 1--5.

\bibitem{constant_envelope}
S.~K. Mohammed and E.~G. Larsson, ``{Constant-Envelope Multi-User Precoding for Frequency-Selective Massive MIMO Systems},'' \emph{IEEE Wireless Communications Letters}, vol.~2, no.~5, pp. 547--550, 2013.

\bibitem{virtualdpd}
C.~Tarver, A.~Balalsoukas-Slimining, C.~Studer, and J.~R. Cavallaro, ``{Virtual DPD Neural Network Predistortion for OFDM-based MU-Massive MIMO},'' in \emph{2021 55th Asilomar Conference on Signals, Systems, and Computers}, 2021, pp. 376--380.

\bibitem{autoprecoder}
X.~Cheng, R.~Zayani, M.~Ferecatu, and N.~Audebert, ``{Efficient Autoprecoder-based deep learning for massive MU-MIMO Downlink under PA Non-Linearities},'' in \emph{2022 IEEE Wireless Communications and Networking Conference (WCNC)}, 2022, pp. 1039--1044.

\bibitem{jointprecodingandpowerallocation}
J.~Jee, G.~Kwon, and H.~Park, ``{Joint Precoding and Power Allocation for Multiuser MIMO System With Nonlinear Power Amplifiers},'' \emph{IEEE Transactions on Vehicular Technology}, vol.~70, no.~9, pp. 8883--8897, 2021.

\bibitem{power_allocation_sofie}
B.~Liu, F.~Rottenberg, and S.~Pollin, ``{Power Allocation for Distributed Massive LoS MIMO with Nonlinear Power Amplifiers},'' in \emph{2022 IEEE 96th Vehicular Technology Conference (VTC2022-Fall)}, 2022, pp. 1--5.

\bibitem{cooperativebeamforming}
J.~Jee, G.~Kwon, and H.~Park, ``{Cooperative Beamforming With Nonlinear Power Amplifiers: A Deep Learning Approach for Distributed Networks},'' \emph{IEEE Transactions on Vehicular Technology}, pp. 1--16, 2022.

\bibitem{z3ro}
F.~Rottenberg, G.~Callebaut, and L.~Van~der Perre, ``{Z3RO Precoder Canceling Nonlinear Power Amplifier Distortion in Large Array Systems},'' in \emph{ICC 2022 - IEEE International Conference on Communications}, 2022, pp. 432--437.

\bibitem{zerofamily}
------, ``{The Z3RO Family of Precoders Cancelling Nonlinear Power Amplification Distortion in Large Array Systems},'' \emph{IEEE Transactions on Wireless Communications}, vol.~22, no.~3, pp. 2036--2047, 2023.

\bibitem{z3ro_val}
T.~Feys, G.~Callebaut, L.~Van~der Perre, and F.~Rottenberg, ``{Measurement-Based Validation of Z3RO Precoder to Prevent Nonlinear Amplifier Distortion in Massive MIMO Systems},'' in \emph{2022 IEEE 95th Vehicular Technology Conference (VTC2022-Spring)}, 2022, pp. 1--5.

\bibitem{dab_mmwave}
S.~R. Aghdam, S.~Jacobsson, and T.~Eriksson, ``{Distortion-Aware Linear Precoding for Millimeter-Wave Multiuser MISO Downlink},'' in \emph{2019 IEEE International Conference on Communications Workshops (ICC Workshops)}, 2019, pp. 1--6.

\bibitem{icc_ccnn}
T.~Feys, X.~Mestre, and F.~Rottenberg, ``{Self-Supervised Learning of Linear Precoders under Non-Linear PA Distortion for Energy-Efficient Massive MIMO Systems},'' in \emph{ICC 2023 - IEEE International Conference on Communications}, 2023.

\bibitem{ofdm}
T.~Hwang, C.~Yang, G.~Wu, S.~Li, and G.~Ye~Li, ``{OFDM and Its Wireless Applications: A Survey},'' \emph{IEEE Transactions on Vehicular Technology}, vol.~58, no.~4, pp. 1673--1694, 2009.

\bibitem{modified_rapp}
C.-S. Choi \emph{et~al.}, ``\BIBforeignlanguage{en}{{RF impairment models for 60GHz-band SYS/PHY simulation}},'' \emph{\BIBforeignlanguage{en}{Project: {IEEE} {P802}.15 {Working} {Group} for {Wireless} {Personal} {Area} {Networks} ({WPANs})}}, p.~17, 2006.

\bibitem{3gpp}
{Nokia}, ``{Realistic power amplifier model for the New Radio evaluation},'' \emph{3GPP TSG-RAN WG4 Meeting 79, R4-163314}, May 2016.

\bibitem{demir2020bussgang}
O.~T. Demir and E.~Bjornson, ``{The Bussgang Decomposition of Nonlinear Systems: Basic Theory and MIMO Extensions [Lecture Notes]},'' \emph{IEEE Signal Processing Magazine}, vol.~38, no.~1, pp. 131--136, 2021.

\bibitem{dist_derivation}
N.~N. Moghadam, G.~Fodor, M.~Bengtsson, and D.~J. Love, ``{On the Energy Efficiency of MIMO Hybrid Beamforming for Millimeter-Wave Systems With Nonlinear Power Amplifiers},'' \emph{IEEE Transactions on Wireless Communications}, vol.~17, no.~11, pp. 7208--7221, 2018.

\bibitem{adam}
\BIBentryALTinterwordspacing
D.~P. Kingma and J.~Ba, ``{Adam: A Method for Stochastic Optimization},'' 2014. [Online]. Available: \url{https://arxiv.org/abs/1412.6980}
\BIBentrySTDinterwordspacing

\bibitem{fundamentals_mimo}
T.~L. Marzetta, E.~G. Larsson, H.~Yang, and H.~Q. Ngo, \emph{{Fundamentals of Massive MIMO}}.\hskip 1em plus 0.5em minus 0.4em\relax Cambridge University Press, 2016.

\bibitem{universalapprox}
K.~Hornik, M.~Stinchcombe, and H.~White, ``\BIBforeignlanguage{eng}{{Multilayer feedforward networks are universal approximators}},'' \emph{\BIBforeignlanguage{eng}{Neural networks}}, vol.~2, no.~5, pp. 359--366, 1989.

\bibitem{inductivebiasold}
\BIBentryALTinterwordspacing
P.~W. Battaglia \emph{et~al.}, ``{Relational inductive biases, deep learning, and graph networks},'' \emph{CoRR}, vol. abs/1806.01261, 2018. [Online]. Available: \url{http://arxiv.org/abs/1806.01261}
\BIBentrySTDinterwordspacing

\bibitem{cnn_gnn}
B.~Zhao, J.~Guo, and C.~Yang, ``{Learning Precoding Policy: CNN or GNN?}'' in \emph{2022 IEEE Wireless Communications and Networking Conference (WCNC)}, 2022, pp. 1027--1032.

\bibitem{graph_rep_learning}
W.~L. Hamilton, ``{Graph Representation Learning},'' \emph{Synthesis Lectures on Artificial Intelligence and Machine Learning}, vol.~14, no.~3, pp. 1--159.

\bibitem{hardware_survey}
\BIBentryALTinterwordspacing
C.~Silvano, D.~Ielmini, F.~Ferrandi, L.~Fiorin, S.~Curzel, L.~Benini, F.~Conti, A.~Garofalo, C.~Zambelli, E.~Calore, S.~F. Schifano, M.~Palesi, G.~Ascia, D.~Patti, S.~Perri, N.~Petra, D.~D. Caro, L.~Lavagno, T.~Urso, V.~Cardellini, G.~C. Cardarilli, and R.~Birke, ``{A Survey on Deep Learning Hardware Accelerators for Heterogeneous HPC Platforms},'' 2023. [Online]. Available: \url{https://arxiv.org/abs/2306.15552}
\BIBentrySTDinterwordspacing

\bibitem{lut_activations}
F.~Piazza, A.~Uncini, and M.~Zenobi, ``{Neural networks with digital LUT activation functions},'' in \emph{Proceedings of 1993 International Conference on Neural Networks (IJCNN-93-Nagoya, Japan)}, vol.~2, 1993, pp. 1401--1404 vol.2.

\bibitem{flops2}
\BIBentryALTinterwordspacing
R.~Hunger, \emph{{Floating Point Operations in Matrix-vector Calculus}}.\hskip 1em plus 0.5em minus 0.4em\relax Munich University of Technology, Inst. for Circuit Theory and Signal Processing, 2005. [Online]. Available: \url{https://books.google.be/books?id=EccIcgAACAAJ}
\BIBentrySTDinterwordspacing

\bibitem{nn_accel}
L.~Baischer, M.~Wess, and N.~TaheriNejad, ``{Learning on Hardware: A Tutorial on Neural Network Accelerators and Co-Processors},'' 2021.

\bibitem{distortion_beamformed2}
C.~Mollen, U.~Gustavsson, T.~Eriksson, and E.~G. Larsson, ``\BIBforeignlanguage{eng}{{Spatial Characteristics of Distortion Radiated From Antenna Arrays With Transceiver Nonlinearities}},'' \emph{\BIBforeignlanguage{eng}{IEEE transactions on wireless communications}}, vol.~17, no.~10, pp. 6663--6679, 2018.

\bibitem{amp_eff}
A.~He, S.~Srikanteswara, K.~K. Bae, T.~R. Newman, J.~H. Reed, W.~H. Tranter, M.~Sajadieh, and M.~Verhelst, ``{Power Consumption Minimization for MIMO Systems — A Cognitive Radio Approach},'' \emph{IEEE Journal on Selected Areas in Communications}, vol.~29, no.~2, pp. 469--479, 2011.

\bibitem{manticore}
F.~Zaruba, F.~Schuiki, and L.~Benini, ``{Manticore: A 4096-Core RISC-V Chiplet Architecture for Ultraefficient Floating-Point Computing},'' \emph{IEEE Micro}, vol.~41, no.~2, pp. 36--42, Mar. 2021.

\end{thebibliography}

\appendices
\section{Polynomial PA parameters}\label{ap:paparams}
\begin{table}[H]
\centering
    \caption[Caption for LOF]{\gls{pa} parameters of the $11^{\mathrm{th}}$ order polynomial PA model at different back-off values ($\beta_1=1$). }
    \label{tab:paparams}
    \centering

 \begin{tabular}{@{}cccccccc@{}} 
 \toprule
 IBO [\SI{}{\decibel}] & $\beta_3\,(\cdot10^{-2})$ & $\beta_5\,(\cdot10^{-3})$ & $\beta_7\,(\cdot10^{-5})$ & $\beta_9\,(\cdot10^{-7})$ & $\beta_{11}\,(\cdot10^{-9})$ \\ [0.5ex] 
     \midrule

 -9  &  \makecell{-4.38184836 \\ -10.1466832j} &  \makecell{1.50490437 \\ 8.422084885j} &  \makecell{-3.13452827\\-28.1868627j} &  \makecell{ 3.49967293\\42.06333106j} &  \makecell{-1.59432984\\-23.1868139j} \\ 
 \arrayrulecolor{black!30}\hline

  -7.5  &  \makecell{-5.79334438 \\ -9.36769411j} &  \makecell{2.39315994 \\ 7.94859107j} &  \makecell{-5.57663136\\-26.92641291j} &  \makecell{6.65066314\\40.4837957j} &  \makecell{-3.14808144\\-22.4280442j} \\ 
\hline
   -6  &  \makecell{-7.50994886 \\ -8.42352484j} &  \makecell{3.66782506 \\ 7.26453523j} &  \makecell{-9.54049052\\-24.8371067j} &  \makecell{12.2703316\\37.5613932j} &  \makecell{-6.13183499\\-20.8924283j} \\ 
\hline
    -4.5  &  \makecell{-9.35828409 \\ -7.41305601j} &  \makecell{ 5.16172165 \\ 6.46522185j} &  \makecell{-14.4481282\\-22.2483069j} &  \makecell{ 19.4963213\\33.7874265j} &  \makecell{-10.0752209\\-18.8479147j} \\ 
\hline
     -3  &  \makecell{-11.1143930 \\ -6.30816977j} &  \makecell{ 6.60156653 \\ 5.47141526j} &  \makecell{-19.1451680\\-18.6610370j} &  \makecell{ 26.2822435\\28.0380833j} &  \makecell{-13.6811147\\-15.4579691j} \\ 
\hline
      -1.5  &  \makecell{-12.903319 \\ -5.49758824j} &  \makecell{ 8.21176444 \\ 4.85204392j} &  \makecell{-24.8588087\\-16.8144990j} &  \makecell{35.2215545\\25.6527492j} &  \makecell{-18.8139985\\-14.3562319j}\\ 
\hline
       0  &  \makecell{-14.4473655 \\ -4.67375592j} &  \makecell{9.58442261 \\ 4.13617338j} &  \makecell{-29.6362436\\-14.3570171j} &  \makecell{42.5309097\\21.9271142j} &  \makecell{-22.9128062\\-12.2805850j} \\ 
       \arrayrulecolor{black}
  \bottomrule
\end{tabular}

\end{table}

\end{document}